\documentclass[a4,11pt]{article}

\usepackage{amsmath,amssymb,color,graphicx,amscd,amsfonts,epsf}

\newcommand\Tab[1]{Table~\ref{tab:#1}}
\newcommand\beq{\begin{eqnarray}}
\newcommand\eeq{\end{eqnarray}} 
\newcommand\Eq[1]{Eq.~\ref{eq:#1}}
\newcommand\Fig[1]{Fig.~\ref{fig:#1}}
\newcommand\Sec[1]{Sec.~\ref{sec:#1}}

\newcommand\bfn{{\bf n}}
\newcommand\bfmu{{\bf \mu}}
\newcommand\bfnu{{\bf \nu}}

\allowdisplaybreaks[4]

\date{}
\begin{document}

\begin{flushright} 
 
KEK-CP-284

\end{flushright} 

\vspace{0.1cm}

\begin{center}
  {\LARGE
  
   
Large-$N_c$ gauge theory and chiral random matrix theory

  }
\end{center}

\vspace{0.2cm}

\begin{center}
         Masanori H{\sc anada}\footnote{
E-mail address : hanada@post.kek.jp}$^{a}$, 		
	Jong-Wan Lee\footnote{
E-mail address : jongwan@post.kek.jp}$^a$  	
and  
	Norikazu Y{\sc amada}\footnote
         {
E-mail address : norikazu.yamada@kek.jp}$^{ab}$

\vspace{0.3cm}

$^a$ {\it KEK Theory Center

 High Energy Accelerator Research Organization (KEK)

		Tsukuba 305-0801, Japan}\\

$^b${\it Department of Particle and Nuclear Physics 

Graduate University for Advanced Studies (SOKENDAI) 

Tsukuba, Ibaraki 305-0801, Japan}\\

\end{center} 


\vspace{1.5cm}

\begin{center}
  {\bf abstract}
\end{center}
  
We discuss how the $1/N_c$ expansion and the chiral random matrix theory ($\chi$RMT) can be used in the study of large-$N_c$ gauge theories.
We first clarify the parameter region in which each of these two approaches is valid: 
while the fermion mass $m$ is fixed in the standard large-$N_c$ arguments ('t Hooft large-$N_c$ limit),
$m$ must be scaled appropriately with a certain negative power of $N_c$
in order for the gauge theories to be described by the $\chi$RMT.
Then, although these two limits are not compatible in general,
we show that the breakdown of chiral symmetry can be detected by combining the large-$N_c$ argument and the $\chi$RMT with some cares.
As a concrete example, we numerically study the four dimensional $SU(N_c)$ gauge theory with $N_f=2$ heavy adjoint fermions, 
introduced as the center symmetry preserver keeping the infrared physics intact, on a $2^4$ lattice. 
By looking at the low-lying eigenvalues of the Dirac operator for a massless probe fermion in the adjoint representation, 
we find that the chiral symmetry is indeed broken with the expected breaking pattern. 
This result reproduces a well-known fact that the chiral symmetry is spontaneously broken 
in the pure $SU(N_c)$ gauge theory in the large-$N_c$ and 
the large-volume limit, and therefore supports the validity of the combined approach. 
We also provide the interpretation of the gap and unexpected $N_c$-scaling, both of which are observed in the Dirac spectrum.

\newpage

\section{Introduction}
\label{sec:introduction}

From a field theoretical point of view, strongly coupled gauge theories
such as quantum chromodynamics (QCD) are of great interest as they
have a number of nontrivial phenomena in themselves.
Because it is difficult to study those theories analytically, effective
theory approaches and the large-$N_c$ limit are often considered.
For example, if one considers the low-energy limit of QCD (and also theories with
spontaneous chiral symmetry breaking (S$\chi$SB)), the form of the low
energy effective theory is tightly constrained by the symmetries so
that one obtains the chiral Lagrangian.
If we further go to the $\epsilon$-regime, where the length of the
box containing the system is much smaller than the pion Compton length,
all such theories fall into one of three {\it universality classes},
which are exactly described by the chiral random matrix theory
($\chi$RMT).
In Ref. \cite{Fukaya:2007fb} this property was used to demonstrate
S$\chi$SB from the first principles by confirming that the spectrum of
the Dirac operator calculated on the lattice agrees with the $\chi$RMT
prediction.
Another example is the 't Hooft large-$N_c$ limit~\cite{'tHooft:1973jz}.
In particular, the large-$N_c$ volume independence (the so-called
Eguchi-Kawai (EK) volume independence) \cite{Eguchi:1982nm} and the
orbifold equivalence \cite{Lovelace:1982hz,Kachru:1998ys} have recently
received large attention in the context of the lattice Monte-Carlo
simulation (see
e.g. \cite{Narayanan:2004cp,Bringoltz:2009kb,Cherman:2010jj}).

Given remarkable successes of these two approaches, it is natural to
consider how they can be combined to study various QCD-like theories.
Once the combined approach has been established, it has several
interesting applications.
One of them is the search for the conformal window or the walking
technicolor model (WTM), where the approach is used to see whether
chiral symmetry is broken or not.
A candidate of the minimal WTM is the $SU(2)$ gauge theory with
two-flavors of fermions in the adjoint representation \cite{Sannino:2004qp, 
Luty:2004ye}.
So far, numerical results from (large-volume) lattice simulations
indicate that this theory is inside the conformal
window~\cite{DelDebbio:2010zz}.
This situation may change when the number of colors increases from
two.
Then, the study of the large-$N_c$ limit provides additional information
useful to understand the phase diagram of the conformal window depicted in
Ref.~\cite{Dietrich:2006cm}.

In this paper, we discuss how to utilize the $\chi$RMT and the
large-$N_c$ equivalence to study the large-$N_c$ gauge theories,
and provide the theoretical argument for the detection of S$\chi$SB 
\footnote{In Ref. \cite{Narayanan:2004cp} and
\cite{Hietanen:2009ex,Hietanen:2012ma}, the authors already used the
$\chi$RMT techniques for numerical studies of chiral symmetry
breaking of $SU(N_c)$ gauge theories in the quenched limit and the
theories with adjoint fermions, respectively.
However, there is a subtlety (the difference of the limiting
procedures explained in \Sec{LargeNcVsRMT}), which makes the conclusion
ambiguous~\cite{Hietanen:2012ma}.
In this paper we clarify this point in order to fully justify the
method.}. 
It turns out that some cares are required because the valid parameter
regions for these two techniques are different in general: 
in the $\chi$RMT limit (see \Sec{LargeNcVsRMT} for the definition), the 't Hooft large-$N_c$ expansion fails due to 
the non-trivial $N_c$-dependence of the expansion coefficients. 
Therefore, we cannot use the EK volume equivalence to relate the small-volume 
with the large-volume lattice theory in the $\chi$RMT limit. 
However, we argue that the chiral properties of the large-volume theory 
can still be studied from the small-lattice model with finite $N_c$ by using 
the analytic continuation provided unbroken center symmetry. 

Bearing the above motivation in mind, as a first step, we study 
pure $SU(N_c)$ gauge theories on a $2^4$ lattice, aiming at
taking the large-$N_c$ limit.
The goal is to find how we should extract physics in the large-volume
limit from small lattice simulations.
We call our theory on a $2^4$ lattice as the $2^4$-lattice model to
avoid a possible confusion with the EK model for which 
the 't Hooft limit is usually assumed. 
Following the argument by Eguchi and Kawai in the 't Hooft limit, we let
our lattice theory keep the center symmetry by introducing two-flavors
of heavy adjoint fermions \footnote{Although we added two heavy adjoint fermions, 
one heavy adjoint fermion is good enough to keep the center symmetry 
\cite{Bringoltz:2009kb,Hietanen:2009ex}}. 
But notice that the theory we deal with is essentially the pure $SU(N_c)$ gauge theory 
since they do not play any role in the low-energy dynamics. 
On the gauge configurations obtained, we calculate the low-lying
spectrum of the Dirac operator in the adjoint representation and compare
with the $\chi$RMT prediction; an agreement between them should provide the
evidence for S$\chi$SB or equivalently non-zero chiral condensate in the 't Hooft limit. 
Using the EK volume equivalence, therefore, we conclude that 
chiral symmetry of pure $SU(N_c)$ gauge theory is indeed broken. 
While the agreement of the Dirac spectrum is confirmed as expected, the
$N_c$-scaling of the spectrum turns out to be different from the one
expected from the usual 't Hooft large-$N_c$ counting. 
To be specific, we found that the eigenvalues scale as $1/N_c$ rather
than $1/N_c^2$ at a reasonably weak coupling. 
We will present a possible explanation for this phenomenon in \Sec{numerical_result}, 
which does not override the occurrence of S$\chi$SB.

This paper is organized as follows.
In \Sec{largeNc}, we review the 't Hooft large-$N_c$ limit and its
properties, especially the EK volume equivalence and the orbifold
equivalence.
In \Sec{RMT} we review the $\chi$RMT - the definition and the
relationship with the $\epsilon$-regime of QCD-like theories with
S$\chi$SB.
In \Sec{LargeNcVsRMT} we discuss the difference of the $\chi$RMT limit
and the 't Hooft limit, in which we can use the $\chi$RMT technique and
the EK volume equivalence, respectively.
Having this difference in mind, we explain how the $\chi$RMT, combined
with numerical simulations on a small lattice, can be used to study
S$\chi$SB of the large-$N_c$ gauge theories.
In \Sec{numerical_result} we present the numerical results of the
$2^4$-lattice model; we first confirm that the center symmetry is
unbroken in the presence of two-flavors of heavy adjoint fermions, and
then proceed to the analysis of the Dirac spectrum including 
comparisons with the $\chi$RMT prediction and determination of the $N_c$-scaling. 

\section{The 't Hooft large-$N_c$ limit}
\label{sec:largeNc}
\begin{figure}
\begin{center}
\includegraphics[width=0.5\textwidth]{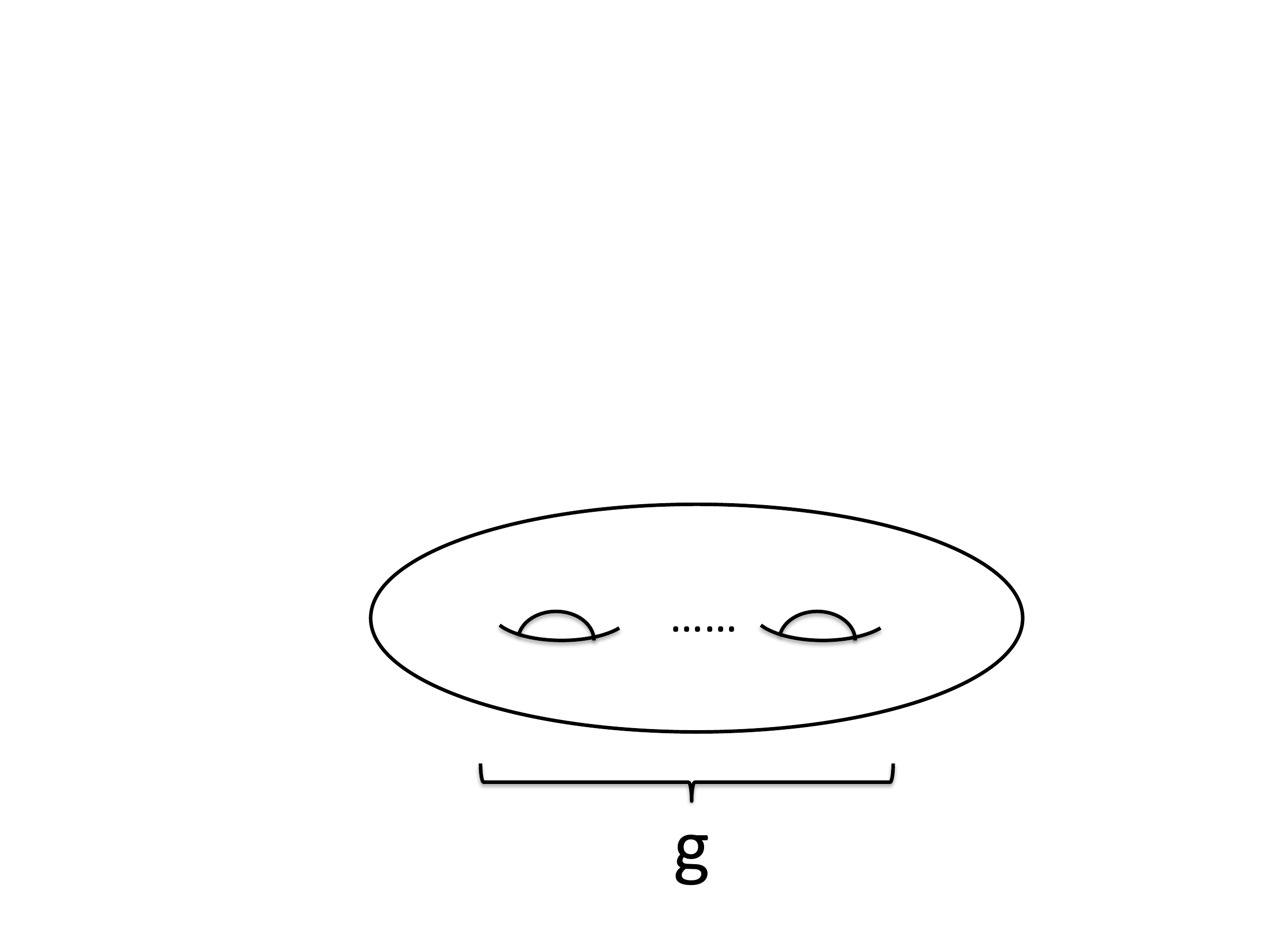}
\caption{
Two-dimensional surface of genus $g$. 
}
\label{fig:genus-g}
\end{center}
\end{figure}

Let us first consider the $SU(N_c)$ pure Yang-Mills theory, 
\begin{eqnarray}
S_{YM}
=
\frac{1}{4g_{YM}^2}\int d^4x\ {\rm Tr} F_{\mu\nu}^2. 
\end{eqnarray}
The 't Hooft large-$N_c$ limit \cite{'tHooft:1973jz} is the large-$N_c$ limit 
in which the 't Hooft coupling constant $\lambda=g_{YM}^2 N_c$ and the space-time $V$ are fixed. \footnote{
More precisely, we take the coupling at some energy scale to be fixed.  
For example, in the case of the lattice regularization, we can take the bare lattice 't Hooft coupling to be the same.  
Then the beta function depends only on $\lambda$ at large-$N_c$, and hence $\lambda$ remains $N_c$-independent 
at any energy scale up to a $1/N_c$-correction.} 
The energy scale under consideration (e.g. distance between operators, the size of the Wilson loop) is also fixed. 
In this limit the $1/N_c$ expansion has a natural 
topological structure, 
\begin{eqnarray}
\langle \hat{O}\rangle = \sum_{g=0}^\infty c_g(\lambda,V) N_c^{-2g},
\end{eqnarray}
where $\hat{O}$ is a properly normalized single trace operator. 
In the perturbation theory, contribution of order $N_c^{-2g}$ comes from the genus-$g$ nonplanar diagrams, 
i.e. the Feynman diagrams which can be drawn  
on the two-dimensional surface with $g$ handles (Fig.~\ref{fig:genus-g}). 
The connected correlation functions of more than one operators have the same structure. 
Therefore the $1/N_c$-expansion is the same as the genus expansion. 
In the string theory, a genus-$g$ surface corresponds to the string world-sheet with $g$ closed string loops. 
Actually the Feynman diagrams can naturally be regarded as dynamical triangulations of the two-dimensional surfaces, 
and $1/N_c^2$ can be identified with the string coupling constant. In Maldacena's gauge/gravity duality conjecture \cite{Maldacena:1997re}, 
certain gauge theories are explicitly related to string theories. 
In the large-$N_c$ limit, only the genus zero diagrams (planar diagrams) survive, or in the string terminology, 
the quantum string effect is suppressed at large-$N_c$. 

Next let us consider QCD with $N_f$ fundamental fermions. 
The fermionic part of the action is given by 
\begin{eqnarray}
S_{fermion}=\sum_{f=1}^{N_f}\int d^4x\ \bar{\psi}^{fund.}_f\left(\gamma^\mu D_\mu+m_f\right)\psi_f^{fund.}. 
\end{eqnarray}
In the 't Hooft limit, in addition to the 't Hooft coupling and the space-time volume, 
the fermion mass $m_f$ must also be fixed. 
Because the fermions have $O(N_f N_c)$ degrees of freedom 
while the gluons have $O(N_c^2)$, every time we replace the gluon loop with the fermion loop we obtain the factor $N_f/N_c$. 
Hence the $1/N_c$-expansion becomes 
\begin{eqnarray}
\langle \hat{O}\rangle = \sum_{g,b=0}^\infty c_{g,b}(\lambda,V,m_f) N_c^{-2g}\cdot (N_f/N_c)^{b},
\label{1/N_expansion_fundamental}
\end{eqnarray}
where $b$ is the number of fermion loops. 
Therefore, when $N_f$ is fixed, diagrams with fermion loops are suppressed. 
(They are not suppressed if one considers the adjoint fermion instead.)  
In the 't Hooft large-$N_c$ limit, the parameters $\lambda$, $V$ and $m_f$ are fixed, because otherwise a nontrivial $N_c$-dependence 
can appear through the coefficient $c_{g,b}(\lambda,V,m_f)$. For example, in QCD, if $m_f$ decreases with $N_c$, 
$N_c$-dependent infrared divergence appears because the pion becomes lighter 
and the standard $1/N_c$-counting in the 't Hooft limit can be destroyed completely. 

In the 't Hooft large-$N_c$ limit, various nice properties hold. 
Below we introduce the EK equivalence and the orbifold equivalence. 

\subsection{The large-$N_c$ equivalences in the 't Hooft limit}
\subsubsection{The Eguchi-Kawai equivalence}\label{sec:EguchiKawai}
Let us consider Wilson's lattice gauge theory on the $L^4$ periodic  lattice, 
\beq
S_{W} = 2 N_c^2 b \sum_\bfn \sum_{\bfmu<\bfnu} \left(1-\frac{1}{N_c} \textrm{Re} \textrm{Tr} P_{\mu\nu} (\bfn)\right),
\label{eq:action_wilson}
\eeq
where $\mu,\nu$ run from 1 to 4, $b=1/\lambda=1/(g_{YM}^2N_c)$, $\bfn$ runs through the $L^4$ lattice,  and the plaquette $P_{\mu\nu}$ is given by 
\begin{eqnarray}
P_{\mu\nu}(\bfn) = U_{\bfn,\mu}U_{\bfn+\mu,\nu}U^\dagger_{\bfn+\nu,\mu}U^\dagger_{\bfn,\nu}. 
\label{eq:plaquette}
\end{eqnarray}

The unitary link variables $U_{\bfn,\mu}$ are related to the Hermitian gauge field $A_\mu$ by $U_{\bfn,\mu}=e^{iaA_\mu(\bfn)}$, 
where $a$ is the lattice spacing. 
The nature of the theory is characterized by the expectation values of the Wilson loops 
\begin{eqnarray}
W(C)={\rm Tr}\left(
U_{\bfn,\mu}U_{\bfn+\mu,\nu}U_{\bfn+\mu+\nu,\rho}\cdots
\right), 
\end{eqnarray}
where $C$ is a closed path on the lattice and the right hand side is the trace of the product of the link variables along $C$. 
Note that the loop $C$ can be larger than the lattice; one can just write a closed loop on the infinite lattice and 
impose the periodic boundary condition. 
This theory has the $({\mathbb Z}_{N_c})^4$ center symmetry, which multiply a phase factor to the link variables: 
\begin{eqnarray}
U_{\bfn,\mu}\to e^{2\pi i k_\mu/N_c} U_{\bfn,\mu}
\qquad
(k_\mu\in{\mathbb Z}).  
\label{eq:center_symmetry}
\end{eqnarray}

Let us consider the 't Hooft large-$N_c$ limit, in which the coupling constant $b$ and the lattice size $L$ are fixed. 
As long as the $({\mathbb Z}_{N_c})^4$ center symmetry is not broken spontaneously, the expectation values of the Wilson loops 
do not depend on $L$. This is called the Eguchi-Kawai equivalence \cite{Eguchi:1982nm}. 
One can also introduce fermions with the periodic boundary condition; for example, for both the fundamental and adjoint fermions, 
the Dirac spectrum does not depend on $L$ when $N_f$ and the mass are fixed. 
Therefore one can study the large-$N_c$ theory on the infinite lattice by using a small lattice, say $1^4$ or $2^4$.    
 
In the pure glue theory, the $({\mathbb Z}_{N_c})^4$ center symmetry is actually broken in the weak coupling limit $b\to\infty$ with fixed $L$ \cite{Bhanot:1982sh} . 
The breakdown of the center symmetry can easily be understood by calculating the one-loop effective action as a function of the Wilson line phases, 
by assuming four Wilson line phases to be static and diagonal.    
Then the off-diagonal fluctuation of  the link variables provides an attractive interaction between the diagonal elements, so that the eigenvalues of 
the link variables favor the same value and hence the center symmetry breaks. 
It is nothing but the deconfinement transition in a very small volume. 
\subsubsection*{Introducing the adjoint matter}

The situation changes drastically when one adds the adjoint fermions, because they provide the repulsive force between eigenvalues. 
When one massless Majorana adjoint fermion is introduced, the theory is roughly the dimensional reduction of the four-dimensional ${\cal N}=1$ pure super Yang-Mills theory.  
In the continuum theory, the forces acting on the Wilson line phases cancel to all order in perturbation theory. 
By taking into account the nonperturbative effects, 
both the center-symmetric \cite{Kovtun:2007py} and center-broken \cite{Hanada:2009hq} phases can exist. 
(The importance of the nonperturbative effect was nicely demonstrated in a related context in \cite{Aoki:1998vn}.)    
Whether the center symmetry breaks or not on a lattice is a subtle issue which depends on the detail of the lattice regularization.  
If we add more massless adjoint fermions, the center symmetry is unbroken irrespectively of the detail of the lattice action \cite{Kovtun:2007py}. 
Furthermore, somehow surprisingly at first glance, the center symmetry does not break spontaneously even with very heavy adjoint fermions, 
whose mass is as heavy as the ultraviolet cutoff scale 
\cite{Bringoltz:2009kb,Azeyanagi:2010ne};  
therefore one can study the pure Yang-Mills theory by using the EK equivalence\footnote{
Another way to avoid the center symmetry breakdown is the double trace deformation \cite{Unsal:2008ch,Vairinhos:2011gv}. 
Other variants, quenched \cite{Bhanot:1982sh} and twisted \cite{GonzalezArroyo:1982hz} EK models, 
turned out to fail actually \cite{Bringoltz:2008av}\cite{Bietenholz:2006cz,Azeyanagi:2007su,Teper:2006sp}. 
For a further modification of the twisted EK model may preserve the center symmetry, see \cite{GonzalezArroyo:2010ss}.}. 

\subsubsection{The orbifold equivalence}\label{sec:orbifold}
Another large-$N_c$ equivalence, which turned out to be deeply related to the EK equivalence, was discovered by Lovelace \cite{Lovelace:1982hz}. 
He considered the $SO(2N_c)$, $USp(2N_c)$ and $SU(N_c)$ Yang-Mills theories and found that the Wilson loops take the same expectation values 
in all three theories. This equivalence was rediscovered and generalized in the study of the string theory \cite{Kachru:1998ys}, 
and a deeper understanding in terms of the field theory was obtained \cite{Bershadsky:1998cb,Kovtun:2004bz}. 
Today it is called the orbifold equivalence. 

The general statement is as follows. Let us start with the `parent' theory, which is $SO(2N_c)$ or $USp(2N_c)$ in the case of \cite{Lovelace:1982hz}. 
We identify a discrete symmetry of the parent theory, which is a ${\mathbb Z}_2$ subgroup of the gauge symmetry in this example. 
Then we perform the `orbifold projection' by keeping only the degrees of freedom which are invariant under the discrete symmetry, 
so that the `daughter' theory ($SU(N_c)$) is obtained. 
If the projection satisfies a certain condition, the correlation functions of the operators in the parent theory which are invariant under the projection symmetry 
(the ${\mathbb Z}_2$ subgroup) 
and the correlation functions of the corresponding operators in the daughter theory take the same values, up to a calculable combinatoric factor. 
In particular, the expectation values of the chiral condensate take the same value in the $SO(2N_c)$, $USp(2N_c)$ and $SU(N_c)$ theories with the fundamental fermions. 
The EK equivalence can also be understood as a special example of the orbifold equivalence \cite{Kovtun:2007py}. 
Other valuable applications include the large-$N_c$ QCD at finite density \cite{Cherman:2010jj,Hanada:2011ju,
Cherman:2011mh,Hidaka:2011jj}, confinement in pure Yang-Mills theory \cite{Unsal:2007vu,Shifman:2008ja}, 
and interesting properties of non-supersymmetric daughters from supersymmetric parents \cite{Schmaltz:1998bg,Strassler:2001fs,Armoni:2003gp}.  

\section{The chiral random matrix theory ($\chi$RMT)}
\label{sec:RMT}
In this section we provide a brief review of the $\chi$RMT. 
For more details, see e.g. \cite{Verbaarschot:2000dy, Akemann:2007rf}.

We consider the $\epsilon$-regime of QCD, 
where the space-time volume $V=L^4$ is taken such that $L$ is much smaller than the pion Compton wavelength 
and is much larger than $1/\Lambda_{QCD}$ \cite{Leutwyler:1992yt}, 
\begin{eqnarray}
\label{eq:low_epsilon}
\frac{1}{\Lambda_{QCD}} \ll L \ll \frac{1}{m_{\pi}}. 
\label{eq:epsilon_regime}
\end{eqnarray}
In the $\epsilon$-regime, the only relevant degrees of freedom are zero momentum modes of pions. 
Then the system has a {\it universality}, that is, the dynamics is determined 
by the symmetry breaking pattern and the microscopic details of the theory do not matter. 
Therefore QCD can be replaced by the $\chi$RMT, which is a random matrix model with the same symmetry breaking pattern; 
in particular, the spectrum of the Dirac operator can be calculated from the $\chi$RMT.  
Note that the same argument holds for other QCD-like  theories as long as chiral symmetry is broken spontaneously. 

The partition function of the $\chi$RMT is given by 
\begin{eqnarray}
\label{eq:RMT}
Z=\int d\Phi \prod_{f=1}^{N_f} \det {\cal D}_f \ e^{-\frac{N \beta}{2}G^2 {\rm tr} \Phi^{\dagger} \Phi},
\end{eqnarray}
where $\Phi$ is a $N \times (N+\nu)$ matrix and $G$ is a normalization parameter. 
$N$ corresponds to the size of the system (roughly speaking the space-time volume), and $\nu$ is the topological charge. 
Correspondingly to the thermodynamic limit of QCD, $N$ is sent to infinity. In this limit, however, the fermion mass $m_f$ must be scaled so that $m_fN$, 
which is (roughly speaking) identified with $m_fV$, is fixed. 

Note that we can define the 't Hooft large-$N$ limit (not the large-$N_c$ limit) for the $\chi$RMT, in which $m_f$ is fixed. 
The limit one has to take for the comparison to QCD ($m_fN$ fixed) is {\it not} this 't Hooft limit.  
This difference is crucial when we compare the large-$N_c$ gauge theories and $\chi$RMT, as we will see in \Sec{LargeNcVsRMT}. 
In addition, it is important to notice that $N$ is identified by the total degrees of freedom associated with the low-energy dynamics 
and the individual degrees of freedom, such as the volume and the number of colors, does not appear in the $\chi$RMT explicitly. 

The ensemble and the Dirac operator ${\cal D}$ are chosen so that the Dirac operator has the same symmetries as the counterparts in QCD and QCD-like theories. 
Depending on the universality classes, there are three $\chi$RMTs, which are distinguished by the Dyson index $\beta=1$, $\beta=2$, and $\beta=4$ \cite{Verbaarschot:1994qf}:
\begin{itemize}
\item{
$\beta=2$ (e.g. QCD and $SU(N_c)$ ($N_c\ge 3)$ with the fundamental fermions):
\begin{eqnarray}
\label{eq:Dyson2}
{\cal D}_f
=
\left(
\begin{array}{cc}
m_f \textbf{1}& \Phi \\
-\Phi^\dagger & m_f \textbf{1}
\end{array}
\right),
\end{eqnarray}
where $\Phi$ is an $ N\times (N + \nu)$ complex matrix and $m_f$ ($f=1,2,\cdots,N_f$)
are the fermion masses.}

\item{
$\beta=1$ (e.g. $SU(2)$ and $USp(2N_c)$ with the fundamental fermions): 
\begin{eqnarray}
\label{eq:Dyson1}
{\cal D}_f
=
\left(
\begin{array}{cc}
m_f \textbf{1}& \Phi \\
-\Phi^T  & m_f \textbf{1}
\end{array}
\right),
\end{eqnarray}
where $\Phi$ is an $ N\times (N + \nu)$ real matrix.}

\item{
$\beta=4$ (e.g. $SU(N_c)$ with the adjoint fermions and $SO(2N_c)$ with the fundamental fermions):
\begin{eqnarray}
\label{eq:Dyson4}
{\cal D}_f
=
\left(
\begin{array}{cc}
m_f \textbf{1} & \Phi \\
-\Phi^\dagger & m_f \textbf{1}
\end{array}
\right),   
\end{eqnarray}
where $\Phi$ is a $2N \times 2(N + \nu)$ quaternion real matrix, 
which can be written by using four $N \times (N + \nu)$ real matrices $\phi_0, \phi_1, \phi_2$ and $\phi_3$ as 
\begin{eqnarray}
\Phi
=
\left(
\begin{array}{cc}
\phi_0+i\phi_3 & i\phi_1+\phi_2\\
 i\phi_1-\phi_2 & \phi_0-i\phi_3. 
\end{array}
\right). 
\end{eqnarray}
} 
\end{itemize}

Because of its simplicity, the $\chi$RMT can be solved analytically. It has been applied not just to the test of S$\chi$SB in the lattice simulations, 
but also to other important problems such as the QCD at a finite baryon chemical potential \cite{Stephanov:1996ki,Osborn:2005ss} and 
the phase structure of the Wilson Dirac operator \cite{Akemann:2010em,Splittorff:2012hz,Akemann:2012bc}.  

\section{Large-$N_c$ versus $\chi$RMT}
\label{sec:LargeNcVsRMT}
In this section, we establish the way to use the $\chi$RMT techniques in large-$N_c$ gauge theories.
For concreteness, we consider the $SU(N_c)$ lattice theory with volume $V$ and fermion mass $m$. 
Here the volume $V$ is arbitrary, although we concentrate on a small fixed volume in Sec.~\ref{sec:numerical_result}. 
\subsection{Difference of the limits}
In order to compare the large-$N_c$ gauge theory and $\chi$RMT, we must understand the difference of the limits, 
which are required for the standard $1/N_c$-counting and the universality, respectively:

\begin{itemize}
\item
When one compares the $\chi$RMT with the gauge theory, the matrix size $N$ of the $\chi$RMT is identified with 
the degrees of freedom in the gauge theory which are important for the low-energy dynamics, 
$N\sim VN_c^\alpha$, where the constant $\alpha$ depends on the representation of the fermion in general. 
(As we will see, $\alpha=1$ for the massless probe adjoint fermion of the $2^4$-lattice model. 
Note that it is different from the usual counting 
in the 't Hooft limit, $\alpha=2$.)
In order for the universality to hold, $mN\sim mVN_c^\alpha$ must be fixed as we take the large-$N_c$ limit. 
Let us call it the `$\chi$RMT limit'. This limit is compatible with the condition for the $\varepsilon$ regime in \Eq{epsilon_regime}.  

\item
For the standard $1/N_c$ counting, the ordinary 't Hooft large-$N_c$ limit, in which $m$ and $V$ are fixed, must be taken. In this limit the large-$N_c$ equivalences (e.g. the EK equivalence)  hold. 
 
\end{itemize}

The $\chi$RMT limit is different from the 't Hooft limit and the standard 't Hooft $1/N_c$-counting does not hold in this limit. 
In order to see this, let us consider the $k$-point connected correlation function in the $\chi$RMT 
(see e.g. \cite{Verbaarschot:1994qf})
\begin{eqnarray}
\langle(\bar{\psi}\psi)^k/N\rangle_{conn,RMT}
=
\left\langle
\frac{1}{N}
\sum_{i=1}^{2N}\left(\frac{1}{\lambda_i+m}\right)^k
\right\rangle_{RMT}, 
\end{eqnarray} 
where $\lambda_i$ are eigenvalues of the Dirac operator. 
In the 't Hooft counting, $\langle(\bar{\psi}\psi)^k/N\rangle_{conn,RMT}$ is of order $N^0$. It is true when $m$ is of order one: 
$1/(\lambda_i+m)^k$ is of order one and the summation over $2N$ order-one quantities is of order $N$. 
However when $m$ scales as $1/N$, the smallest of $\lambda_i+m$ is also of order $1/N$, and hence 
the correlation function becomes of order $N^{k-1}$. This divergence is analogous to the infrared divergence 
in $SU(3)$ QCD in the $\varepsilon$-regime. 
(Note that $\bar{\psi}\psi$ in $\chi$RMT corresponds to $\int d^4x\bar{\psi}\psi$ in QCD.) 
In large-$N_c$ field theories, this corresponds to the divergence with a certain power of $N_c$, 
which is different from the 't Hooft counting.  
This peculiar behavior can also be understood in terms of the large-$N_c$ gauge theory; 
because the coefficients $c_{g,b}$ in \eqref{1/N_expansion_fundamental} are functions of $m$ and $V$, 
they can have nontrivial $N_c$-dependences in the $\chi$RMT limit, where $m$ and $V$ are scaled with $N_c$. 

Because of this difference, the large-$N_c$ equivalences do not hold in the $\chi$RMT limit. 
As an example, let us consider the chiral condensate in $SU(N_c)$, $SO(2N_c)$ and $USp(2N_c)$ theories. 
In the planar limit, because of the orbifold equivalence (Sec.~\ref{sec:orbifold}), they are the same as a function of $m$ \cite{Cherman:2010jj,Hanada:2011ju}. 
On the other hand, in the $\chi$RMT limit they can be calculated by using the $\chi$RMTs 
as a function of $mN$, and they do behave differently \cite{Verbaarschot:1995yi,Toublan:1999hi}. 
It means the orbifold equivalence does not hold there. 
(Note that it is the case even in the quenched theory, which we numerically study in this paper.) 

\subsection{The comparison}
First let us recall how one can realize S$\chi$SB in numerical
simulations of QCD in the $\varepsilon$-regime.
The criterion for S$\chi$SB is whether the Dirac spectrum agrees with
the $\chi$RMT prediction.
Once they are found to agree, the value of the chiral condensate is
determined.
Notice that the value of the chiral condensate obtained in the
$\varepsilon$-regime is the same as the one obtained by taking the
chiral limit after the large-volume (or thermodynamic) limit.
It means that, whatever the order of the limiting procedures is, we can
determine the chiral condensate.

\begin{figure}
\begin{center} 
\includegraphics[width=.5\textwidth]{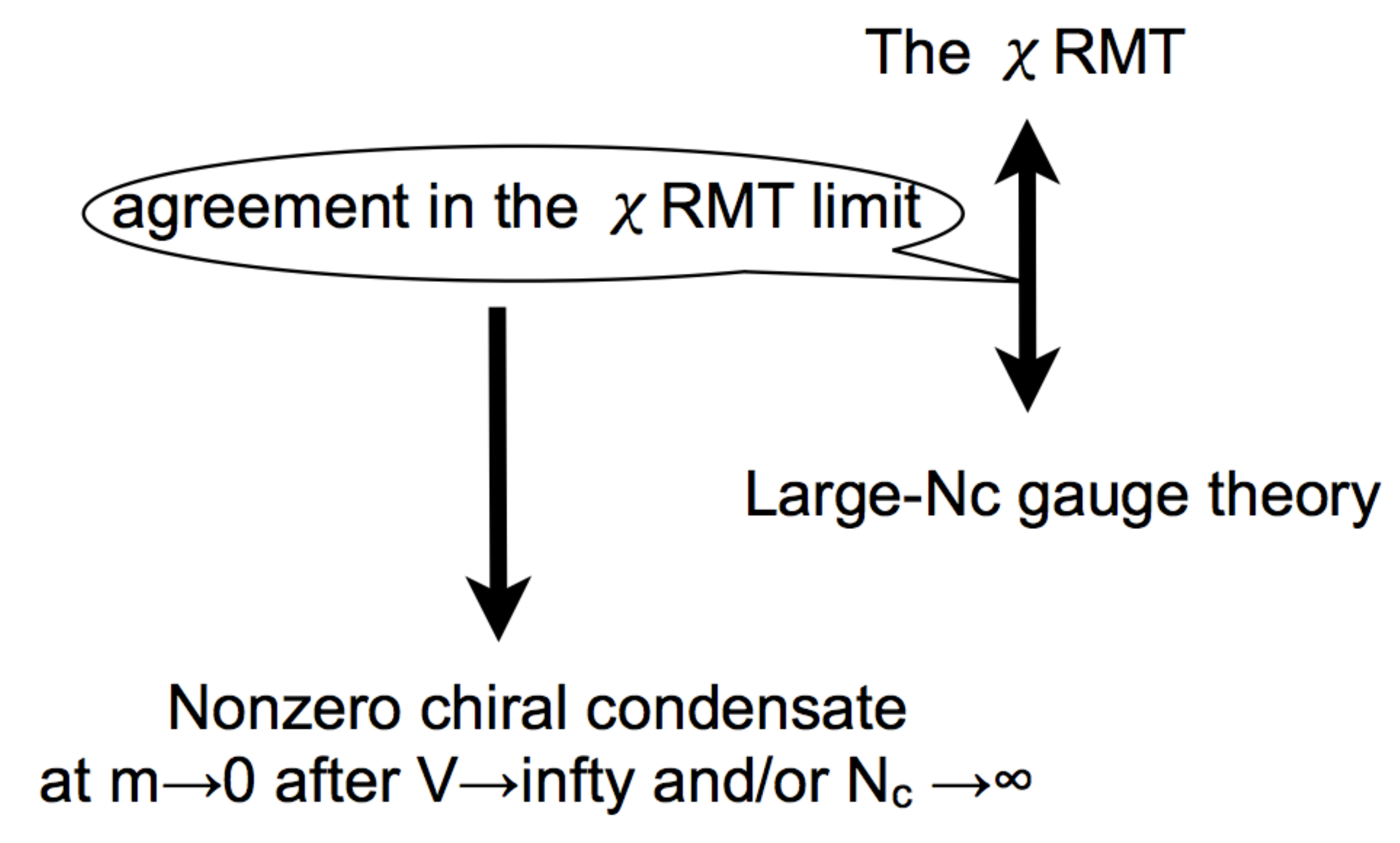}  
\end{center}
\caption{
 The agreement between the $\chi$RMT and the large-$N_c$ gauge theory in
 the $\chi$RMT limit assures S$\chi$SB in the 't Hooft limit and/or the
 large-volume limit. 
}
\label{fig:equivalence}
\end{figure}

The same logic should hold for large $N_c$ gauge theories.
But this time, the role of the volume in the $\chi$RMT limit can be
played by $N_c$.
Namely, the $\chi$RMT limit is achieved by taking the large $N_c$ limit
with $mVN_c^\alpha$ fixed. 
Then, the criterion for S$\chi$SB is again the agreement of the Dirac
spectrum in the $\chi$RMT (\Fig{equivalence}).
%

\begin{figure}
\begin{center} 
\includegraphics[width=.6\textwidth]{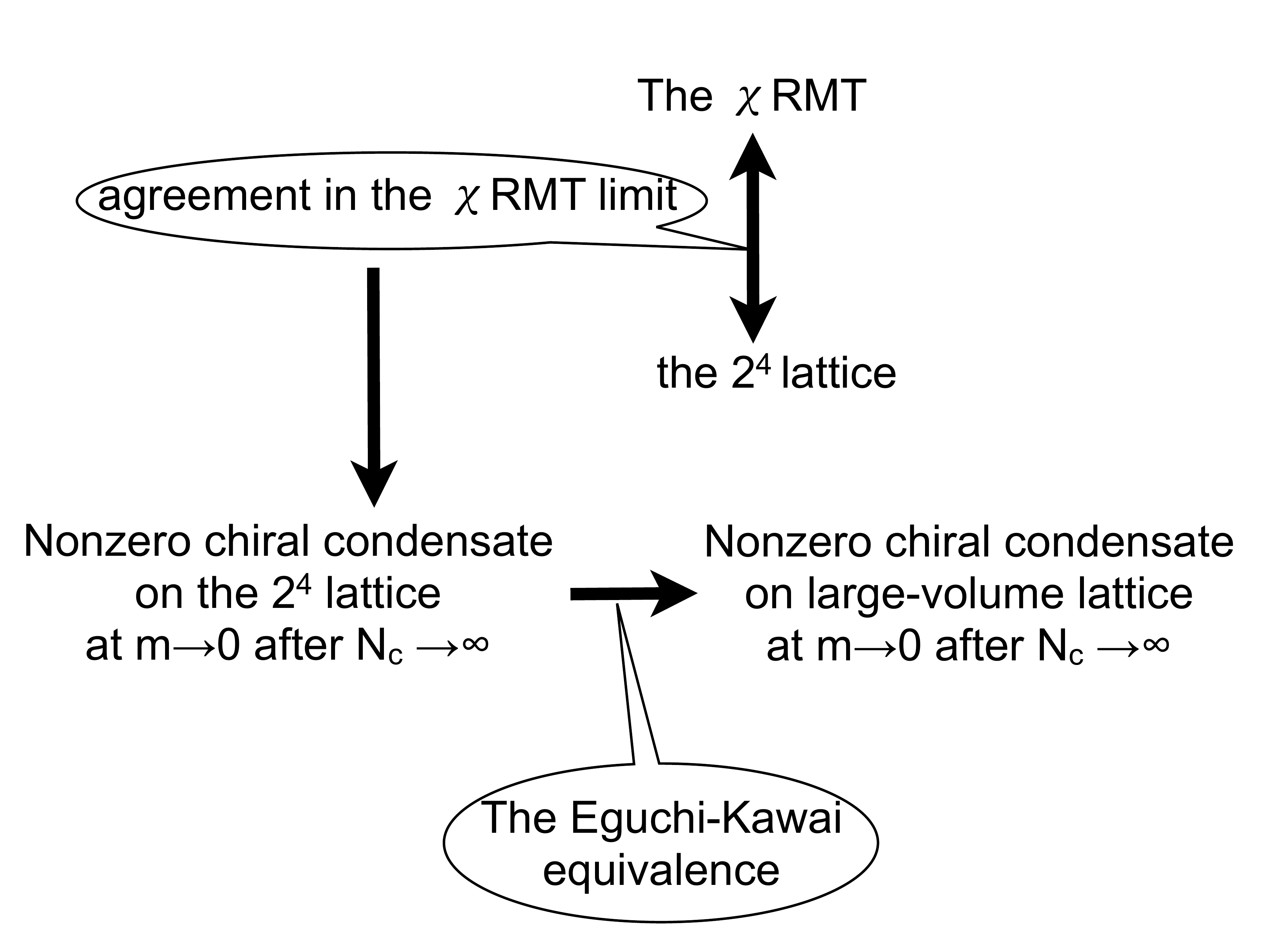}  
\end{center}
\caption{
 The agreement between the $\chi$RMT and the $2^4$-lattice model in the
 $\chi$RMT limit implies S$\chi$SB or equivalently non-zero chiral condensate in the 't Hooft limit.
 Then, one can use the EK equivalence to conclude S$\chi$SB in the
 large-volume theory. 
}
\label{fig:web_of_equivalences}
\end{figure}

In Fig.~\ref{fig:web_of_equivalences} we show how one can combine the
$\chi$RMT and the EK equivalence to see S$\chi$SB.
We compare the spectrum of the Dirac operator of the probe fermion with
the mass $m\sim N_c^{-\alpha}$ in the $V=2^4$ lattice and that in the
$\chi$RMT; the agreement between them implies the breakdown of 
chiral symmetry (nonzero chiral condensate) in the 't Hooft limit.
Using the EK equivalence, therefore, we conclude that chiral symmetry is
spontaneously broken in the large-volume lattice gauge theories.
The constant $\alpha$ is determined by the {\it effective} degrees of
freedom in the low mode region.
As we will see, in the $2^4$-lattice model with the massless probe
adjoint fermion, small eigenvalues scales as $1/N_c$.
Therefore we take $\alpha=1$ \footnote{
In \cite{Hietanen:2012ma} the same scaling has been already found
for the $N_f=1$ theory with a very small mass (which is essentially
massless).
In that paper, however, the authors regard that the deviation from the
't Hooft counting $\alpha=2$ suggests the absence of S$\chi$SB.
}.
We also found a nice agreement of the low-lying Dirac spectrum, which we
interpret as the presence of S$\chi$SB.

Here let us discuss the $N_c$ scaling of the chiral condensate for later use.
In the large-volume theory, the chiral condensate $\Sigma$ of the
adjoint fermion is related with the spectral density $\rho(\lambda)$ by
the Banks-Casher relation
\cite{Banks:1979yr},
\beq
  \Sigma
= |\langle \bar{\psi} \psi \rangle|
= \pi \rho(0),~~\textrm{with}~~\rho(\lambda)
= \left\langle \frac{1}{V}\sum_n \delta(\lambda-\lambda_n)\right\rangle,
\label{eq:BanksCasher}
\eeq
where $\lambda_n$ is the Dirac eigenvalues.
The spectral density at $\lambda=0$ is proportinal to the inverse of
spacing of the near-zero Dirac eigenvalues,
$\Delta \lambda=\lambda_{i+1} - \lambda_i$.
$\Delta \lambda$ is expected to scale like $\sim 1/N$, where $N$ is the
number of degrees of freedom important for the low-energy dynamics. 
Since the definition of $\rho(\lambda)$ in (\ref{eq:BanksCasher}) is already
normalized by $V$, $\rho$ thus defined scales as $N_c$ increases. 
In the 't Hooft limit, the degrees of freedom of both gauge and fermion parts are $O(N_c^2)$  
and as a consequence $\Sigma = N_c^2 \tilde\Sigma$ in the 't Hooft limit, 
where an $O(N^0)$ quantity $\tilde\Sigma$ is defined by the {\it properly normalized operator} as
\beq
  \tilde{\Sigma}  
= \frac{1}{N_c^2} |\langle \textrm{tr} (\bar{\psi} \psi) \rangle|
= \pi \tilde{\rho}(0),~~\textrm{with}~~\tilde{\rho}(\lambda)
=\left\langle \frac{1}{V N_c^2}\sum_n \delta(\lambda-\lambda_n)
 \right\rangle.
\label{eq:largeNcBanksCasher}
\eeq
In the $\chi$RMT limit, however, 
we have to carefully count the number of degrees of freedom associated with 
the low-energy dynamics which may be different from that in the 't Hooft limit; 
in general the properly normalized spectral density would be related with the spacing of 
small Dira eigenvalues by $\tilde{\rho}(0)=1/(VN^\alpha\Delta\lambda)$, where the $\alpha$ 
mentioned above is determined from how the low-lying eigenvalues scale with $N_c$. 

\begin{figure}
\begin{center} 
\includegraphics[width=.6\textwidth]{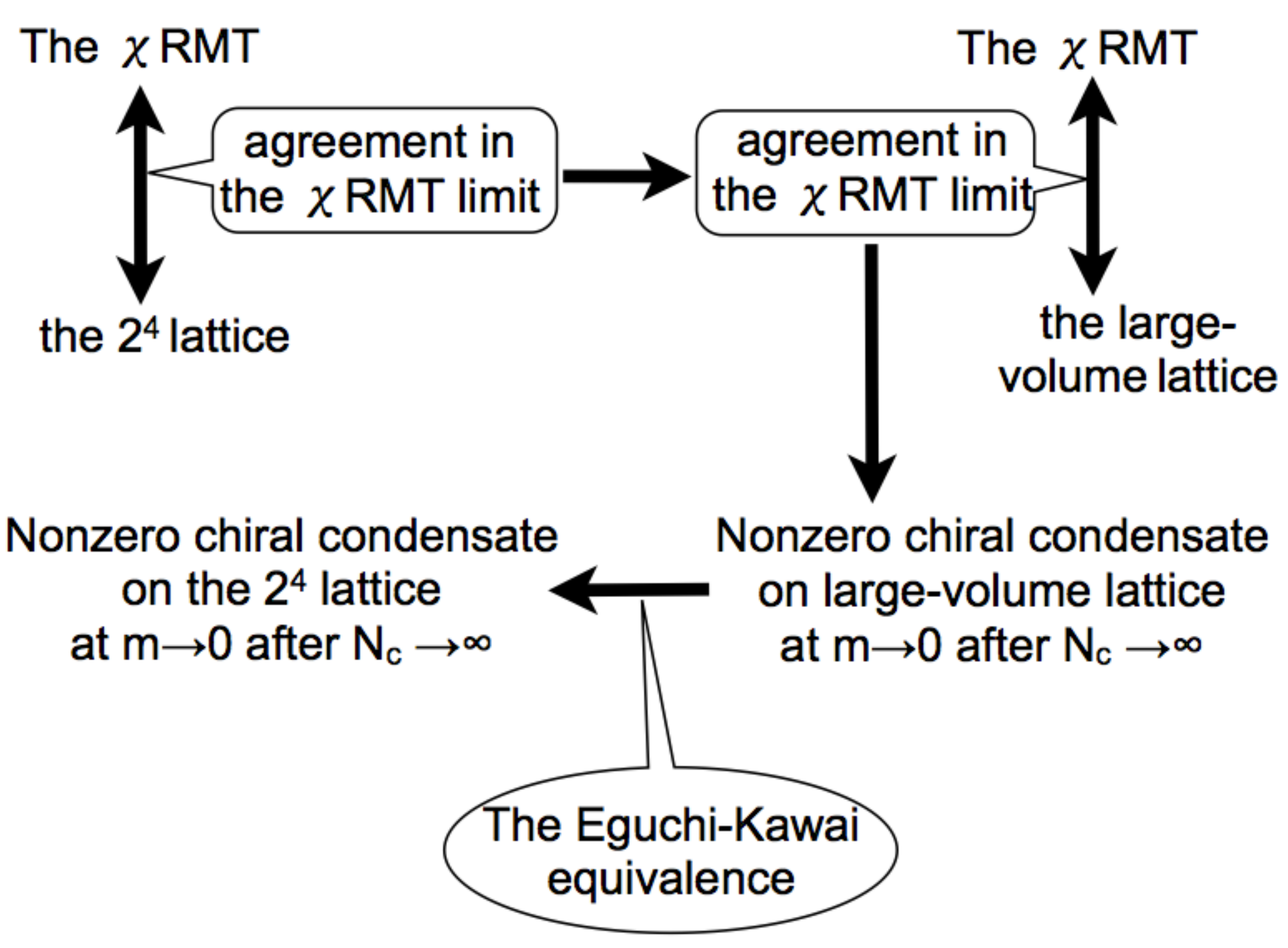}  
\end{center}
\caption{
Another interpretation. The agreement between the $\chi$RMT and the $2^4$-lattice model in the $\chi$RMT limit 
translates into the agreement the $\chi$RMT and the large-volume lattice thanks to the absence of the phase transition (the center symmetry breakdown). 
One can conclude S$\chi$SB of the large-volume theory from the agreement between $\chi$RMT and the $2^4$-lattice model in the $\chi$RMT limit. 
}
\label{fig:web_of_equivalences_2}
\end{figure}

Strictly speaking, there is a subtlety in the $\chi$RMT limit of the $2^4$-lattice model: 
because the space-time volume of this lattice is very small,   
usual derivation of the $\chi$RMT from QCD 
via the chiral perturbation theory may not be applicable\footnote{
One might think the standard mapping rule between the planar sector of the small-volume (i.e. EK model) and the large-volume theory can be used. 
However nonplanar diagrams can contribute in the $\chi$RMT limit.  
}. In order to circumvent this subtlety, one can take another path (\Fig{web_of_equivalences_2}) as follows. 
First let us consider a sufficiently large volume, where the usual derivation of the $\chi$RMT is valid 
when chiral symmetry is spontaneously broken. Then, the distribution of the Dirac eigenvalues should agree with 
that from the $\chi$RMT after properly normalizing the eigenvalues by $\lambda (VN^\alpha)\tilde{\Sigma}$. 
Therefore, as long as $V$ is large enough to justify the chiral perturbation, 
$V$-dependence does not appear manifestly. Now let us shrink the volume further. 
If we consider the pure Yang-Mills without the (heavy) adjoint fermion, there is a phase transition 
(the breakdown of the center symmetry), beyond which one cannot expect the same expression for the distribution of the Dirac eigenvalues. 
However in the present case, because there is no phase transition thanks to the heavy adjoint fermion,  
it is expected that the same expression for the eigenvalue distribution holds at very small volume 
as a function of $\lambda (VN_c^\alpha) \tilde{\Sigma}$, 
where the value of $\alpha$ could be different from that in the large-volume theory.  
(In order to confirm this argument a quantitative study of $\tilde{\Sigma}$ on a large lattice is required and therefore we leave it 
as our future work.) 
Then, the agreement between the $\chi$RMT and the $2^4$-lattice model in the RMT limit 
translates into the agreement between the $\chi$RMT and the large-volume lattice 
provided unbroken centery symmetry. 
At large volume and in the 't Hooft limit, one can conclude S$\chi$SB in the ordinary manner. 
By further using the EK equivalence, S$\chi$SB of the $2^4$-lattice model in the 't Hooft limit can also be concluded.

\section{Numerical simulation of the $2^4$-lattice model and comparison to $\chi$RMT}
\label{sec:numerical_result}
In this section, we apply the strategy explained in \Sec{LargeNcVsRMT}
to the $2^4$-lattice model and detect S$\chi$SB.
After introducing the lattice action and simulation details in
\Sec{simulation_details}, in \Sec{center_sym} we confirm that the center
symmetry, which is crucial for the use of the EK equivalence,
can be kept unbroken by adding heavy adjoint fermions.
Then, in \Sec{dirac_spectrum} we calculate the Dirac spectrum and
compare with the $\chi$RMT prediction.

For an earlier work along the same direction, see
\cite{Hietanen:2009ex}.
See also \cite{Bringoltz:2011by,GonzalezArroyo:2012st} for recent
numerical simulations of a single-site model.
\subsection{Lattice action and simulation details}
\label{sec:simulation_details}
We take the standard the Wilson plaquette gauge action
\Eq{action_wilson}, 
\beq
  S_g 
= 2 N_c^2 b \sum_\bfn \sum_{\bfmu<\bfnu}
  \left(1-\frac{1}{N_c} \textrm{Re} \textrm{Tr} P_{\mu\nu}
  (\bfn)\right),
  ~~~\bfn \in 2^4.
\eeq
In order to preserve the center symmetry, we add two-flavor of heavy
adjoint fermions, for which we choose the plain Wilson fermion, 
\beq
  S_f 
= \sum_{j=1}^2 \sum_\bfn
   \overline{\psi}_{\bfn,j}\left(\psi_{\bfn,j}
  -\kappa \sum_\mu^4 \left[(1-\gamma_\mu)U_{\bfn,\mu}^{adj} \psi_{\bfn+\mu,j}
  +(1+\gamma_\mu)U_{\bfn-\mu,\mu}^{\dagger,adj} \psi_{\bfn-\mu,j}\right]\right),
\eeq
where $b$ and $\kappa$ represent the inverse of the 't Hooft coupling,
$b=1/(g_{YM}^2 N_c)$, and the hopping parameter,
$\kappa=(2m_0 a+8)^{-1}$ (where $m_0$ is the bare fermion mass),
respectively.
The plaquette $P_{\mu\nu}$ is defined by \Eq{plaquette}.
For the fermionic action, the link variables in the adjoint
representation are defined by
\beq
U_{a,b}^{adj}=\frac{1}{2}\textrm{Tr}[T^a_F U T^b_F U^\dagger],
\eeq
where $T^a_F$ are $SU(N_c)$ generators in the fundamental representation.
This action is invariant under $SU(N_c)$ local gauge transformation 
\beq
U_{\bfn,\bfmu} \longrightarrow \Omega_{\bfn} U_{\bfn,\bfmu} \Omega_{\bfn+\bfmu}^\dagger, ~~~~~\Omega_{\bfn} \in SU(N_c),
\eeq
as well as $({\mathbb Z}_{N_c})^4$ global center transformation
\Eq{center_symmetry}.
Throughout this work, we impose periodic boundary conditions in all
directions for both link variables and fermion fields.

Our lattice simulations consist of two parts: 1) quenched calculations
($\kappa=0$), where $m_0 a$ is infinite, as a nontrivial check of our
numerical code by confirming the breaking of center symmetry at weak
coupling, 2) simulations with two heavy adjoint fermions ($\kappa=0.09$), 
where the center symmetry is unbroken even at weak coupling while 
the low-energy dynamics are essentially the same as those of pure $SU(N_c)$ gauge theory 
since $m_0$ is of order of the inverse lattice spacing. 

Simulation parameters are summarized in \Tab{parameters}.
We performed simulations at $b=0.5$ (weak coupling) for up to $N_c=16$,
which is relatively smaller than that in previous single-site model
simulations \cite{Bringoltz:2009kb,Azeyanagi:2010ne,Bringoltz:2011by}.
As we will see in \Sec{heavy_adjoint_result}, however, we could obtain good
large-$N_c$ limits since we have additional suppression of the finite
volume effects thanks to the larger volume of a $2^4$ lattice.
For $N_c=8$ and $\kappa=0$, we also performed simulations at $b=0.3$ and
$0.4$ corresponding to the strong and intermediate couplings,
respectively.
In addition, we performed simulations at $b=0.2$ (strong coupling) for
$\kappa=0.09$ and $N_c=6,\ 8$.
We used the Hybrid Monte Carlo (HMC) algorithm for all lattice
simulations with the plain leap-flog integrator, where the step size is
tuned such that the acceptance ratio is in the range of 70 -- 80 \%.
The simulation codes were developed from the one unsed in
\cite{Aoki:2008tq} with appropriate modification to $SU(N_c)$ adjoint fermions 
with arbitrary large $N_c$. 
After $200$ trajectories for thermalization, $N_{conf}=138-600$
configurations are accumulated for each ensemble, where every two
adjacent configurations are separated by $10$ trajectories.
Statistical errors are calculated by using the standard bootstrapping technique. 

\begin{table}
\caption{%
\label{tab:parameters}%
Simulation parameters.
}
\begin{center}
\begin{tabular}{cccc||cccc}
\hline \hline
~~~$\kappa$~~~ & ~~~$b$~~~ & ~~~$N_c$~~~ & ~~~$N_{conf}$~~~ & ~~~$\kappa$~~~ & ~~~$b$~~~ & ~~~$N_c$~~~ & ~~~$N_{conf}$~~~ \\
\hline
0 & 0.3 & 8 & 500 & 0.09 & 0.2 & 6 & 465 \\ \cline{2-4}
 & 0.4 & 8 & 500 & &  & 8 & 458\\ \cline{2-4} \cline{6-8}
 & 0.5 & 2 & 200 & & 0.5  & 2 & 150\\
 &  & 3 & 200 &  & & 3 & 150 \\
 &  & 4 & 600 &  & & 4 & 150 \\
 &  & 5 & 500 &  & & 5 & 500 \\
 &  & 6 & 200 &  & & 6 & 500 \\
 &  & 8 & 500 &  & & 8 & 500 \\
 &  & 10 & 500 &  & & 10 & 300 \\
 &  & 11 & 300 &  & & 11 & 400 \\
 &  & 15 & 200 &  & & 12 & 250 \\
 &  & 16 & 400 &  & & 15 & 138 \\
 &  &  &  &  & & 16 & 500 \\
\end{tabular}
\end{center}
\end{table}

\subsection{$({\mathbb Z}_{N_c})^4$ Center Symmetry}
\label{sec:center_sym}
As explained in \Sec{EguchiKawai}, the non-trivial condition for 
the large-$N_c$ EK equivalence is that the
$({\mathbb Z}_{N_c})^4$ center symmetry of the reduced volume theory
must be unbroken. 
The presence of the center symmetry is established by checking the
followings:
(1) the Polyakov loop radially scatters in the vicinity of origin in the
complex plane,
(2) the magnitude of the Polyakov loop approaches zero as $N_c$
increases,
(3) the average plaquette value measured in the reduced model agrees
with that in the large-volume lattice gauge theory.
In this section, we present our findings for pure $SU(N_c)$ gauge theory
and the theory with two heavy adjoint fermions.

\subsubsection{Pure $SU(N_c)$ gauge theory}
\begin{figure}
\includegraphics[width=.33\textwidth]{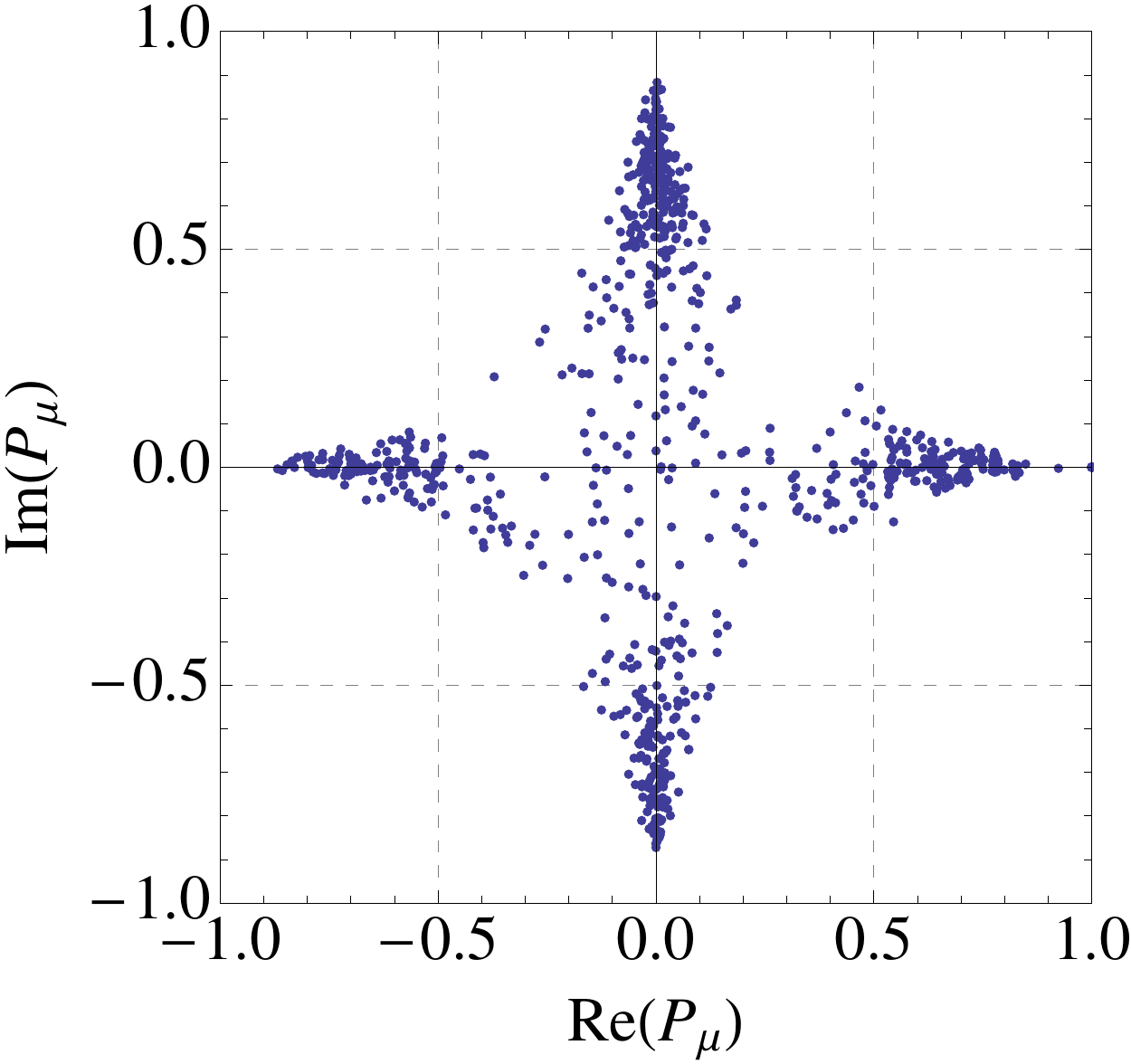}
\includegraphics[width=.33\textwidth]{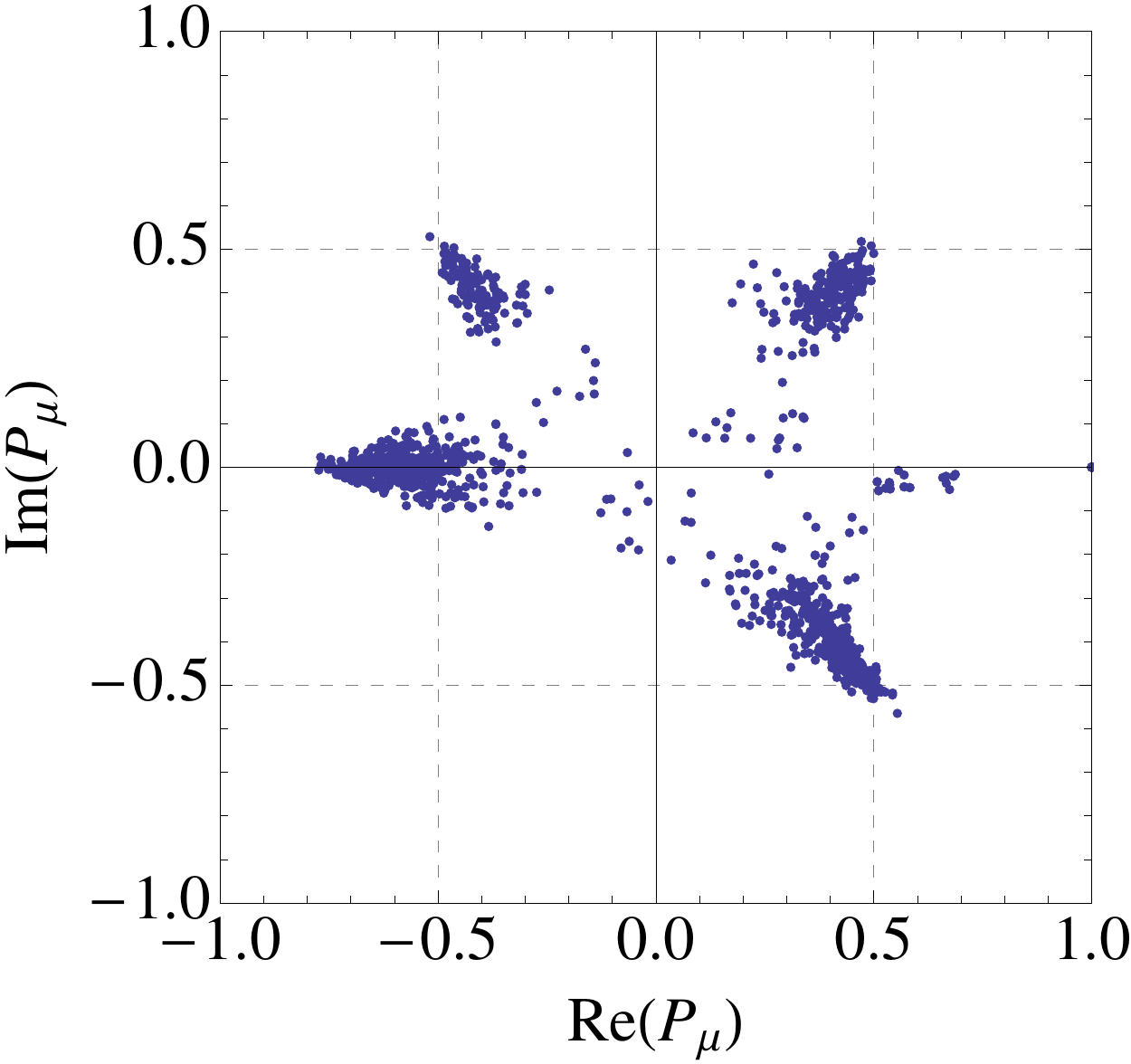}
\includegraphics[width=.33\textwidth]{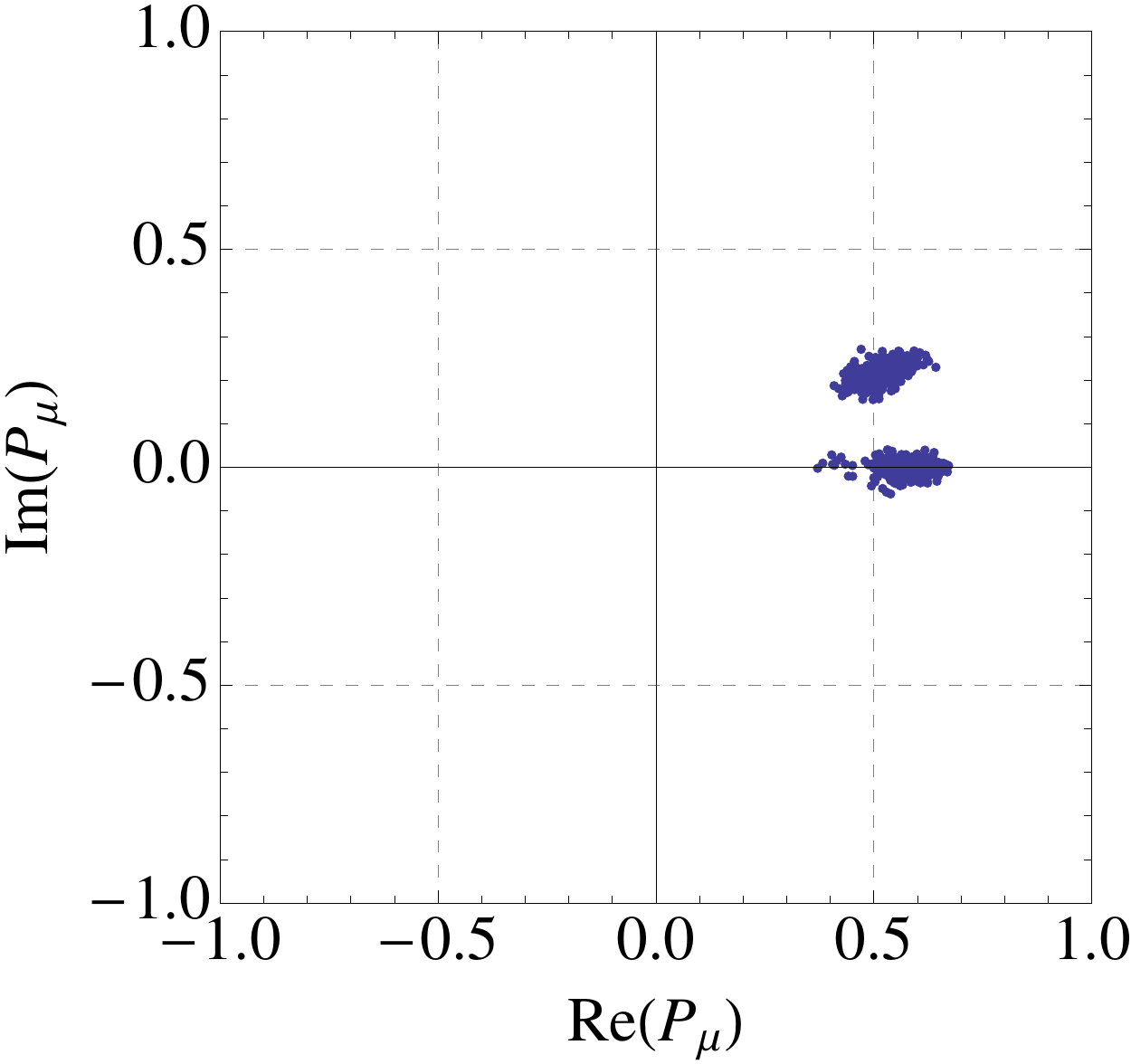}
\caption{%
\label{fig:scatter_Nf0_b0p5}%
Scatter plots of the Polyakov loops for quenched simulations with
 $b=0.5$. The number of colors are $N_c=4, 8$, and $16$ from the left to
 right.
}
\end{figure}

In \Fig{scatter_Nf0_b0p5}, we show scatter plots \footnote{For all
scatter plots, We used about a hundred ensembles and chose one of the
direction $\mu$ out of four directions. We found similar scatter plots
for three other directions.} of the Polyakov loops defined by
\begin{eqnarray}
P_{\mu=1}(y,z,w)
=
\frac{1}{N_c} 
\textrm{Tr} U_{\mu=1;0,y,z,w}U_{\mu=1;1,y,z,w}, 
\end{eqnarray}
and similarly for $\mu=2,3$ and $4$, where $b=0.5$ and $N_c=4, 8, 16$
from left to right.
In this weak coupling regime, the plots clearly show the center symmetry
breaking as the Polyakov loops are localized at the elements of the
center of $SU(N_c)$, $2 i n_\mu \pi/N_c$, where
$n_\mu=0,1,\cdots,N_c-1$.
For a given number of configurations, the number of clusters decreases
to one as the number of colors $N_c$ increases from $5$ to $16$; this
means the tunneling transitions between different center-symmetry-broken
vacua will eventually disappear at $N_c\rightarrow\infty$.
\begin{figure}
\begin{center}
\includegraphics[width=.33\textwidth]{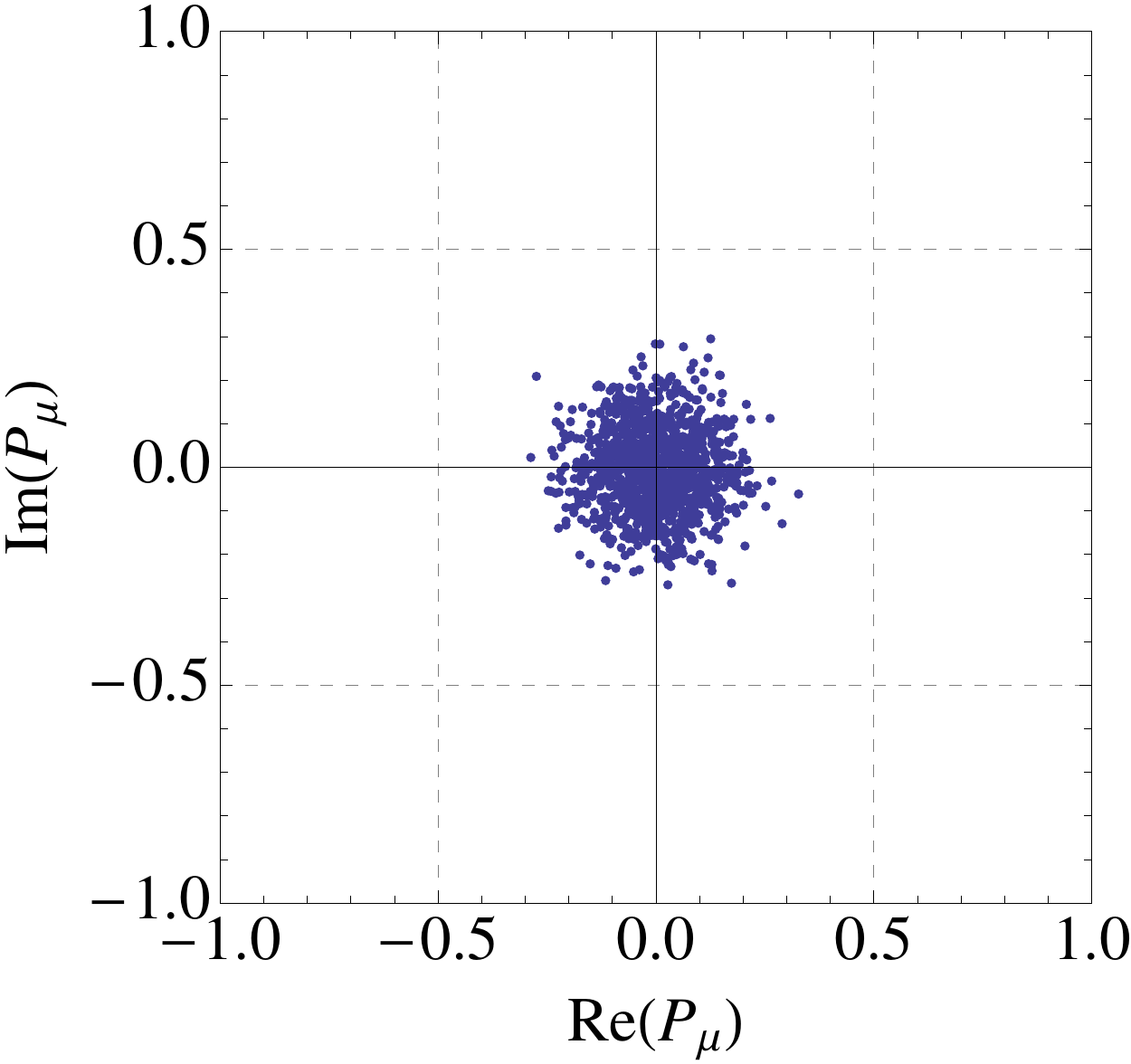}
\hskip .7in
\includegraphics[width=.33\textwidth]{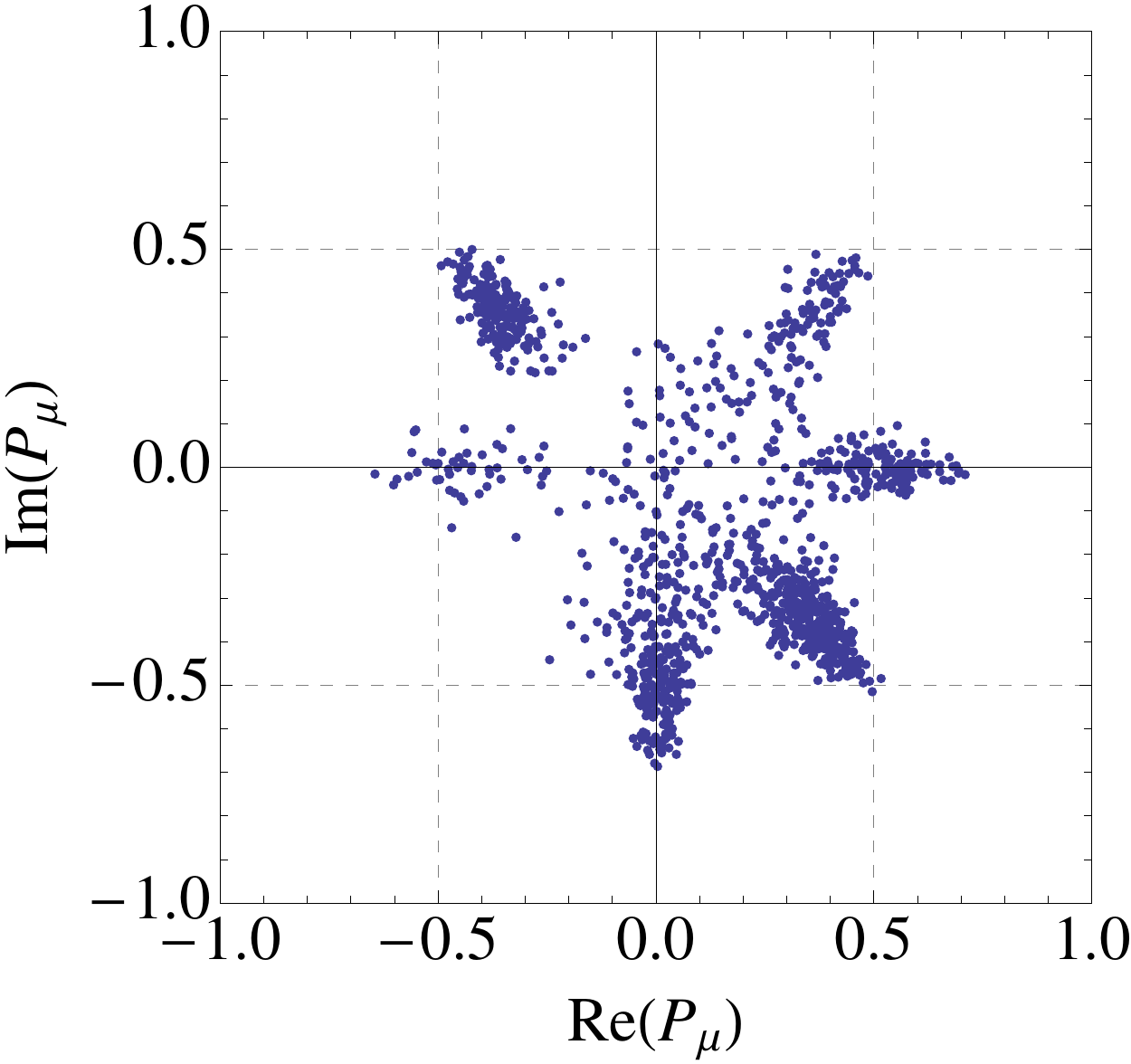}
\caption{%
\label{fig:scatter_Nf0_Nc8}%
 Scatter plots of the Polyakov loops for quenched simulations with
 $N_c=8$.
 The values of the couplings are $b=0.3$ and $0.4$ from left to right.
}
\end{center}
\end{figure}

For $N_c=8$, we performed two more simulations with smaller values of
$b$.
The results of the Polyakov loops are shown in \Fig{scatter_Nf0_Nc8}.
At $b=0.3$, the Polyakov loops develop a cluster around the origin,
while at $b=0.4$ they spread out and are localized at the
center-symmetry-broken vacua like at $b=0.5$.
Therefore, we conclude that the center symmetry is restored in the
strong coupling regime.
The boundary between the strong and weak coupling regimes is located
somewhere between $b=0.3$ and $b=0.4$, which is consistent with the
results in \cite{Catterall:2010gx}.
\begin{figure}
\begin{center}
\includegraphics[width=.5\textwidth]{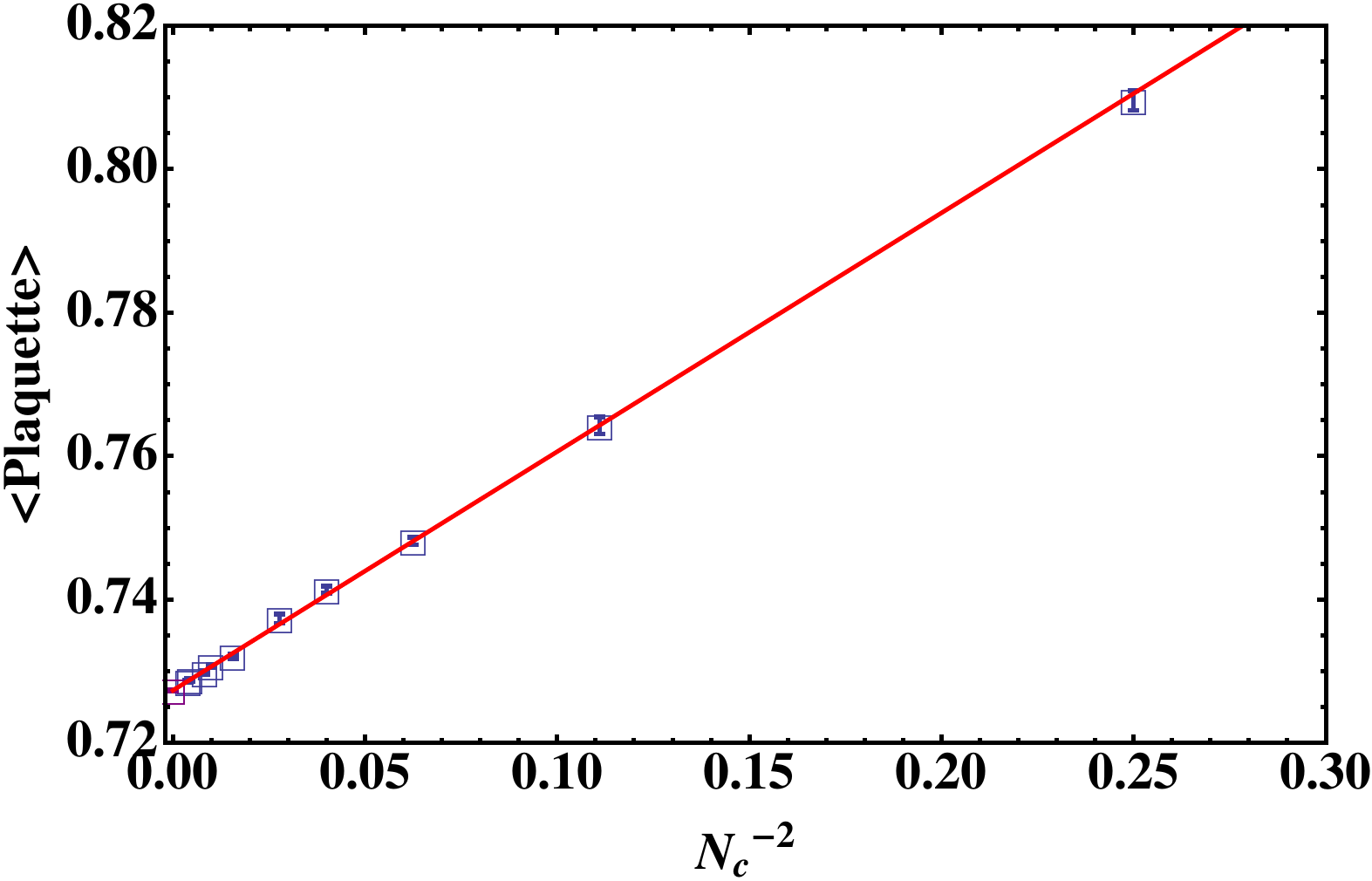}
\caption{%
\label{fig:plaquette_Nf0}%
Average plaquette values for quenched simulations with $b=0.5$. 
 The red solid line represents the uncorrelated fit of the data to the
 function, $c_0+c_1/N_c^2$, and we obtained $c_0=0.72733(12)$ and
 $c_1=0.3325(46)$, where the chi-square/d.o.f is $0.67$.
}
\end{center}
\end{figure}

In \Fig{plaquette_Nf0}, we plotted average plaquette values for $N_c$ up
to $16$.
The measured plaquette values turn out to scale as $1/N_c^2$ and thus we
perform a fit to the data using a constant plus quadratic function of
$N_c^{-1}$ (red solid line in the figure).
We obtained $0.72733 (12)$ in the large $N_c$ limit, which is larger
than $0.7182$, the value from large-volume lattice gauge theory
\cite{Bringoltz:2009kb}.
Therefore we reproduced the well-known fact that the large-$N_c$ volume
reduction for pure Yang-Mills theory fails at weak coupling due to
absence of the center symmetry.

\subsubsection{$SU(N_c)$ gauge theory with two heavy adjoint fermions}
\label{sec:heavy_adjoint_result}
\begin{figure}
\includegraphics[width=.33\textwidth]{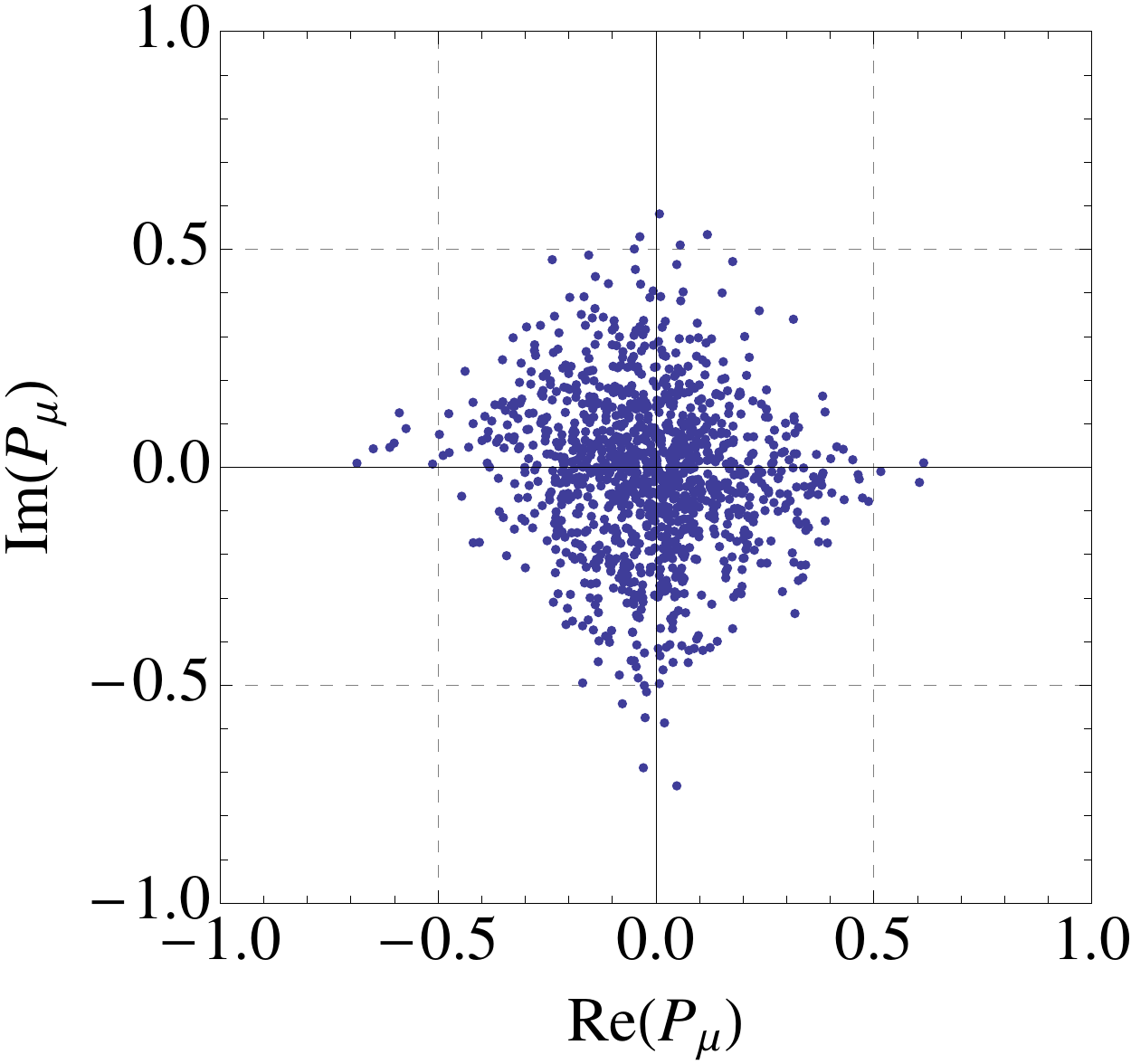}
\includegraphics[width=.33\textwidth]{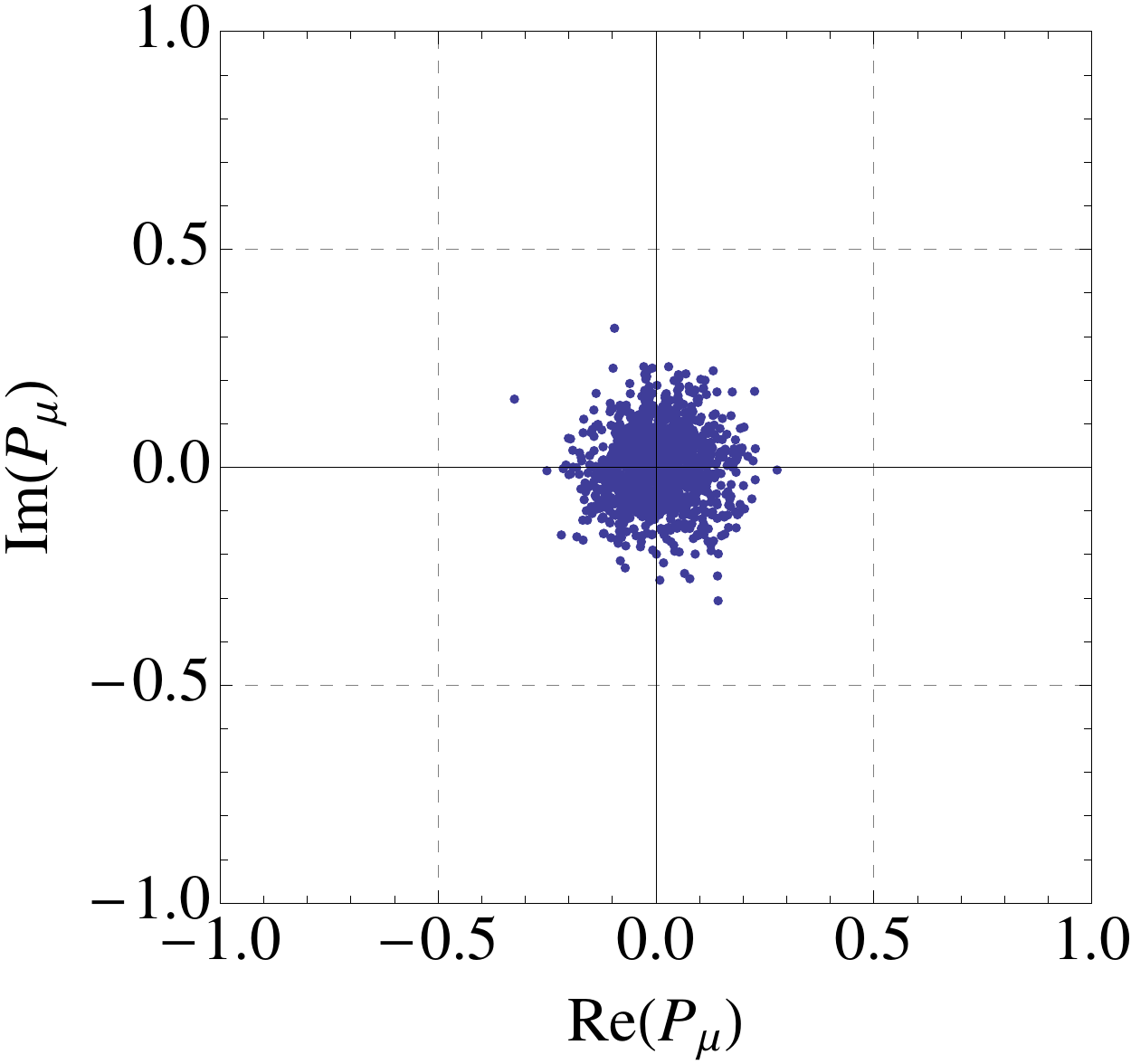}
\includegraphics[width=.33\textwidth]{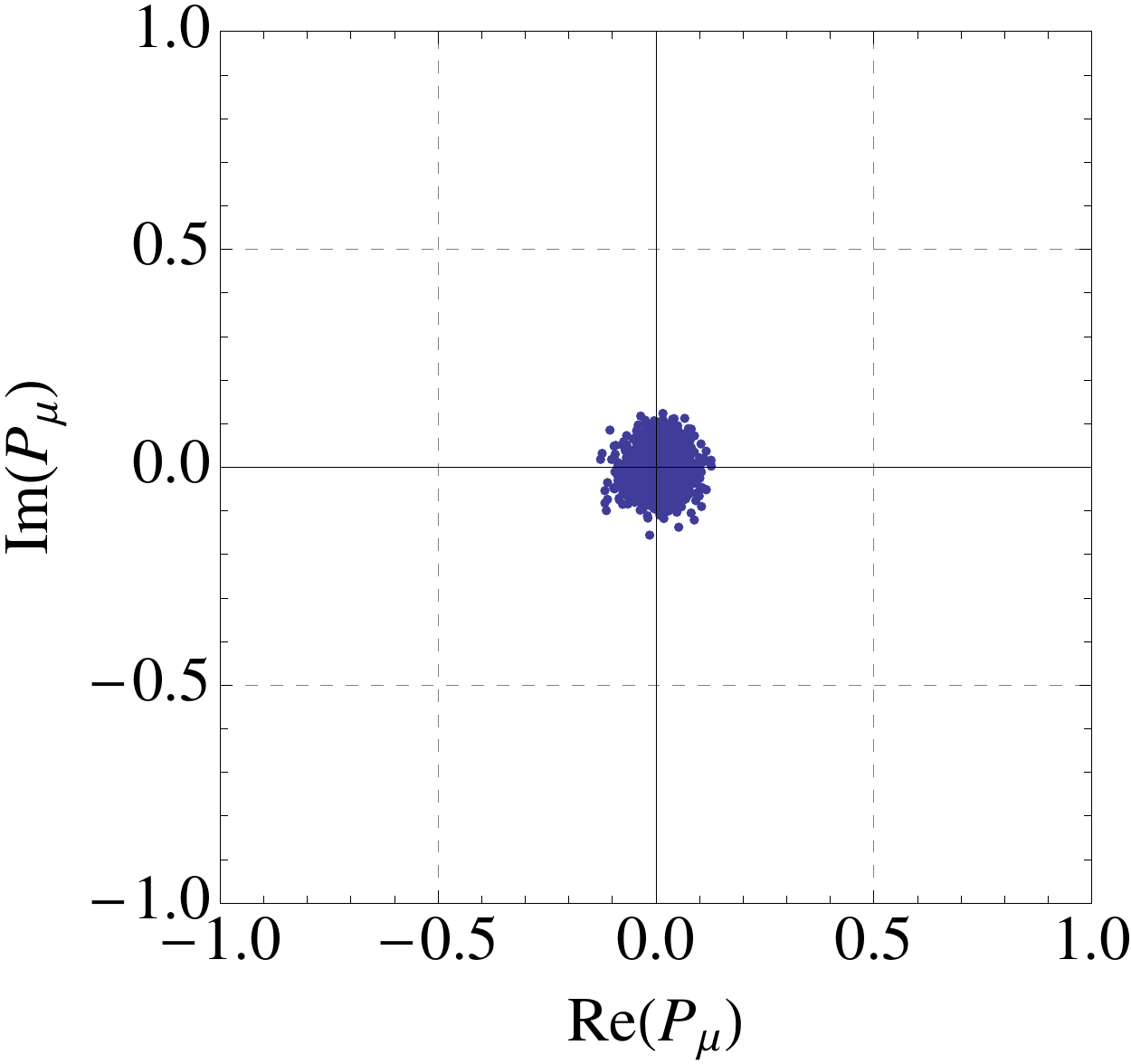}
\caption{%
\label{fig:scatter_Nf2_b0p5}%
 Scatter plots of the Polyakov loops at $b=0.5$ with two heavy adjoint
 fermions.
 The number of colors are $N_c=4, 8$, and $16$ from left to right.
}
\end{figure}

Now let us add two heavy adjoint fermions in order to keep the center
symmetry unbroken even at weak coupling.
At $b=0.5$, we performed a set of simulations for various values of
$N_c$ with fixed value of $\kappa=0.09$ as shown in \Tab{parameters}.
In \Fig{scatter_Nf2_b0p5}, we show scatter plots of the Polyakov loops
for $N_c=4,8,$ and $16$.
The clustering of the Polyakov loops around the origin for $N_c=8$ and
$16$ clearly shows that the center symmetry is intact, which is in
contrast to the case without adjoint fermions.

\begin{figure}
\includegraphics[width=.5\textwidth]{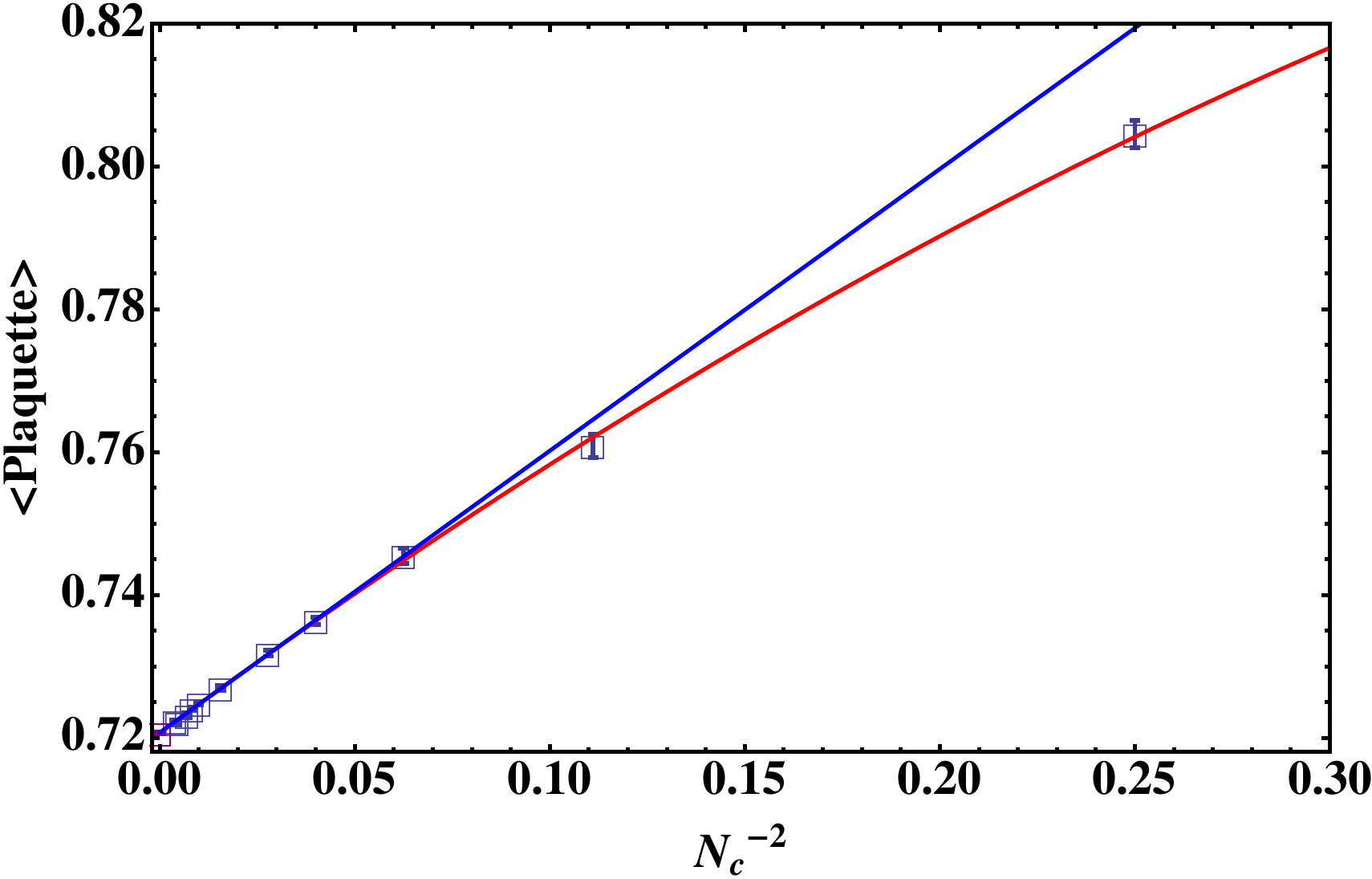}
\includegraphics[width=.5\textwidth]{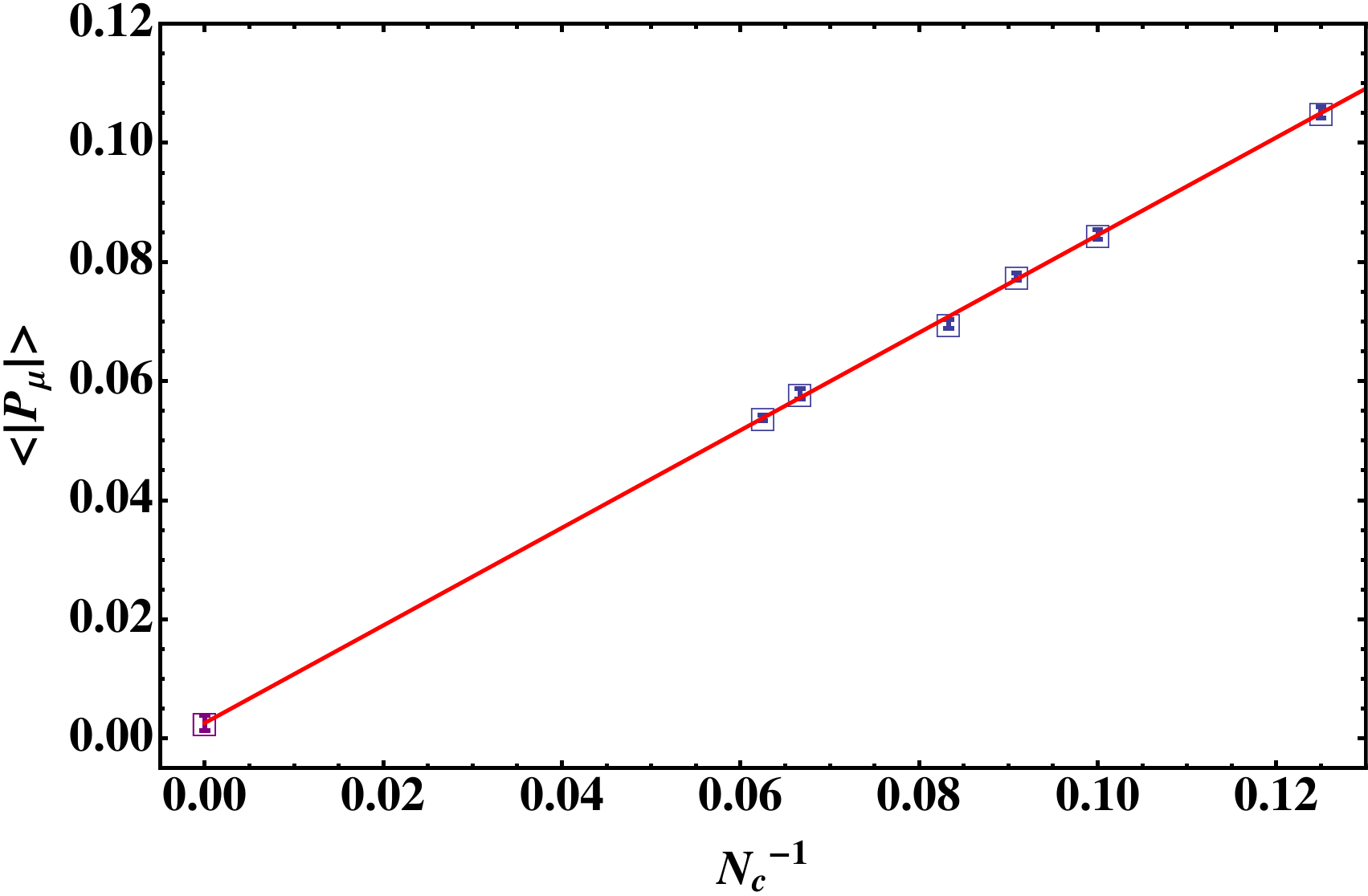}
\caption{%
\label{fig:polyakov_plaquette_Nf2}%
 (Left) Average plaquette values and (Right) average values of the
 magnitudes of the Polyakov loops along with fit results in
 \Tab{polyakov_plaquette_Nf2_tab}.
}
\end{figure}

\begin{table}
\caption{%
\label{tab:polyakov_plaquette_Nf2_tab}%
Fit results for the Polyakov loop and plaquette values.
}
\begin{tabular}{ccccccc}\hline\hline
& Data set & Fit function & $c_0$  & $c_1$ & $c_2$ & $\chi^2$/d.o.f\\\hline
Plaquette & $N_c=[4,16]$ & $c_0+c_1 /N_c+c_2 /N_c^2$ & $0.72053(71)$ & $0.003(13)$ & $0.383 (50)$& $0.47$\\
& $N_c=[2,16]$ & $c_0+c_1 /N_c^2+c_2 /N_c^4$ & $0.72067(14)$ & $0.405(12)$ & $-0.283(61)$ & $0.49$\\\hline
Polyakov loop & $N_c=[5,16]$ & $c_0+c_1 /N_c$ & $0.0026(12)$ & $0.819(15)$ & & $0.89$\\\hline\hline
\end{tabular}
\end{table}

In \Fig{polyakov_plaquette_Nf2}, we show the average plaquette values
versus $N_c^{-2}$ and the magnitudes of the Polyakov loops versus
$N_c^{-1}$.
For the  plaquette values, we examined two different fits, one fitting
the data of $4\leq N_c \leq 16$ to $c_0+c_1/N_c+c_2/N_c^2$ (blue solid
line) and the other fitting those of $2\leq N_c \leq 16$ to
$c_0+c_1/N_c^2+c_2/N_c^4$ (red solid line).
For the Polyakov loops, we fit the data of $5\leq N_c \leq 16$ to
$c_0+c_1/N_c$ to obtain the large $N_c$ limit.
The fit results are summarized in \Tab{polyakov_plaquette_Nf2_tab}.

As discussed in \cite{Azeyanagi:2010ne} and \cite{Bringoltz:2011by} in
great details, the leading correction to plaquette in the large-$N_c$
limit for a single-site lattice model is $O(1/N_c)$, instead of
$O(1/N_c^2)$ (in the usual 't Hooft counting), due to the contributions
of diagonal zero modes.
If we apply this argument to our non-single site reduced model, the
$1/N_c$ correction term is suppressed by $1/V$.
Indeed, we obtained a consistent result where the one-sixteenth
of the coefficient of $1/N_c$ calculated in \cite{Azeyanagi:2010ne}
\footnote{The value of the coefficient is $0.22\pm 0.01$, which was not
presented in the original paper.} is within the uncertainty of our
results shown in \Tab{polyakov_plaquette_Nf2_tab} (first row).
The extracted plaquette value also agrees with that obtained in a
single-site model \cite{Azeyanagi:2010ne}, but it is systematically
larger than that from the large-volume lattice calculation for pure
Yang-Mills.
This tiny difference comes from the presence of heavy fermions.
The magnitude of the Polyakov loop goes to zero as $N_c$ increases; it
scales as $1/N_c$ in the asymptotic region.
Therefore, we conclude that the center symmetry is unbroken for a given
lattice parameters and thus the EK volume equivalence is applicable.

\subsection{Comparison to $\chi$RMT}
\label{sec:dirac_spectrum}
As discussed in \Sec{LargeNcVsRMT}, to detect S$\chi$SB we compare the
low-lying Dirac eigenmodes of the $2^4$-lattice model with those of
$\chi$RMT in the limit of $N_c \rightarrow \infty$ with $mVN_c^\alpha$
fixed.
The simplest way to achieve the $\chi$RMT limit without losing
the generality might be taking $m=0$ and $N_c\rightarrow
\infty$\footnote{Although this limit looks like the 't Hooft limit ($m$
is fixed), actually one should not regard it so because the infrared
(IR) regulator is assumed in the usual 't Hooft counting; for the usual
't Hooft counting one must introduce nonzero fixed $m$ as an IR
regulator and takes the large-$N_c$ limit, and then sends the mass to
zero.}. For this purpose, we calculate the low-lying spectrum of the
overlap-Dirac operator $\mathcal{D}$ for a massless fermion in the
adjoint representation. The operator $\mathcal{D}$ is defined by
\beq
  \mathcal{D}
= M\left[1+\gamma_5~\frac{H_w(-M)}{\sqrt{H_w(-M) H_w(-M)^\dagger}}\right],
\eeq
where $H_w(-M)$ is the hermitian Wilson-Dirac operator and the parameter
$M$ is taken to be $1.6$ in the most cases. 
The overlap-Dirac eigenvalues $\tilde{\lambda}_k$ lie on a circle in
the complex plane \cite{Neuberger:1997fp, Neuberger:1998wv}.
To compare the Dirac spectrum with the $\chi$RMT, we consider the
projection of the eigenvalues to the imaginary axis,
\beq
\lambda_k = \frac{\textrm{Im} [\tilde{\lambda}_k]}{1-\textrm{Re} [\tilde{\lambda}_k]/m_0}.
\eeq
Note that the Dirac eigenvalues for adjoint fermions are appearing as
conjugate pairs with 2-fold degeneracy and we take the distinct
eigenvalues on the upper-half plane for our numerical results.
Unless otherwise noted, the numerical results in this section are
restricted to the case of $b=0.5$ (weak coupling).

\subsubsection{Chiral symmetry breaking}
\label{sec:chi_sym_breaking}

\begin{figure}
\includegraphics[width=.5\textwidth]{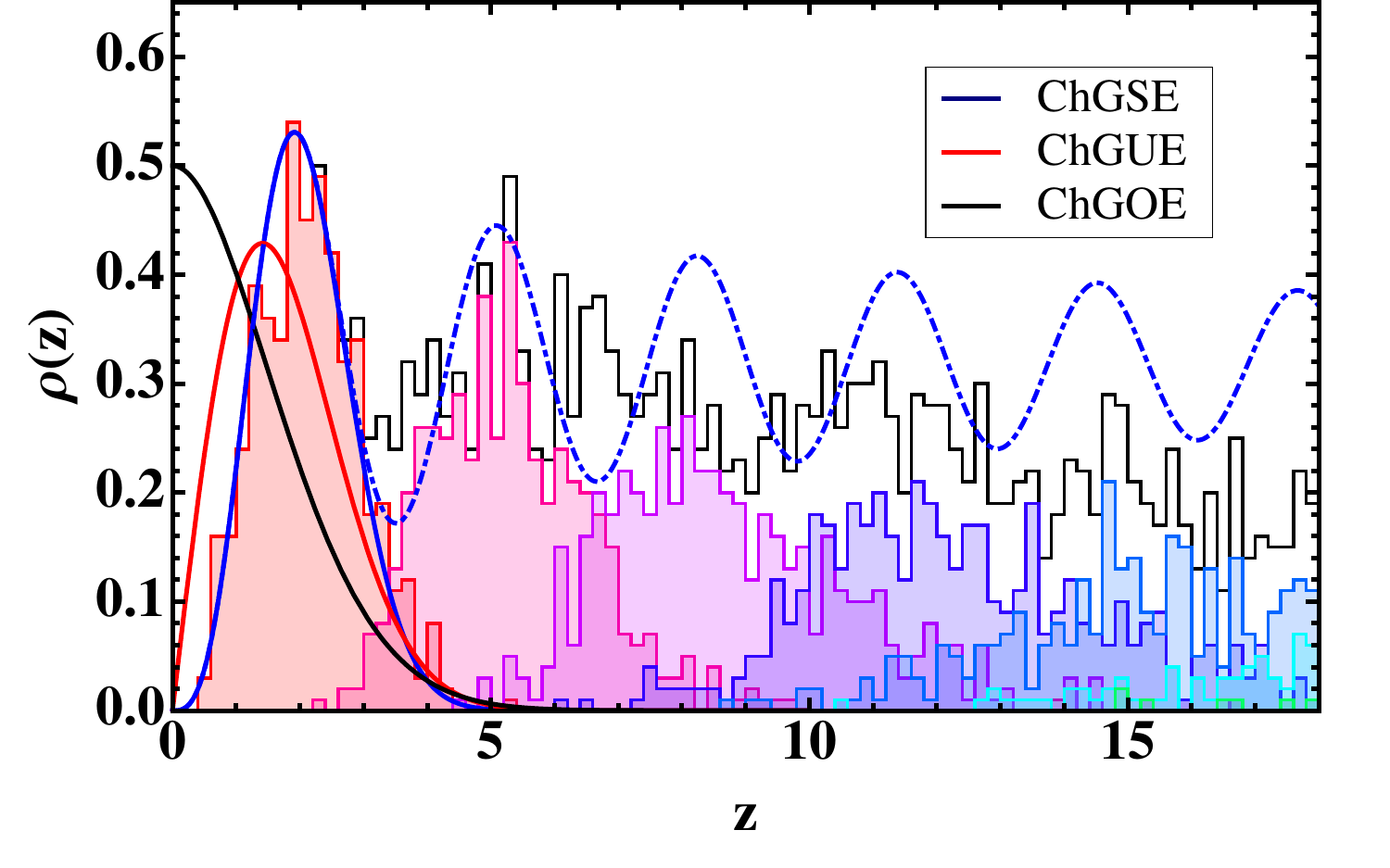}
\includegraphics[width=.5\textwidth]{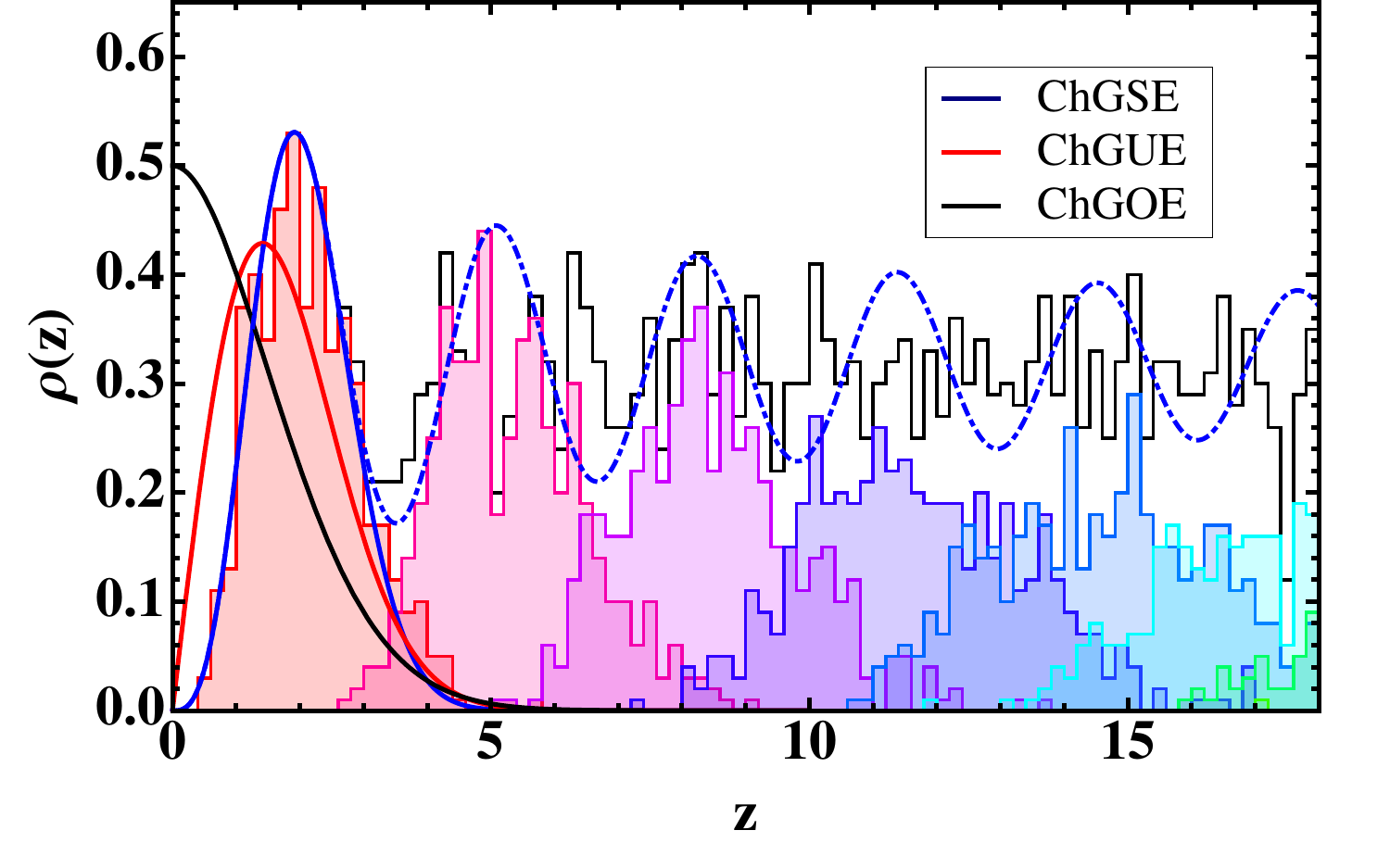}
\caption{%
\label{fig:Dirac_spectrum}%
 Low-lying Dirac spectrum for $N_c=8$ (left) and $N_c=16$ (right).
 The Dirac eigenvalues are rescaled by matching the ensemble average of
 the lowest eigenvalue with the expectation value of the lowest
 eigenvalue in the $\chi$RMT for the ChGSE. The solid and dotted lines
 represent the $\chi$RMT predictions of the distributions of the lowest
 eigenvalue and the spectral density, respectively. The colored
 histograms are for individual eigenvalues, while the black histograms
 are for all eigenvalues.
}
\end{figure}

As seen in \Sec{center_sym}, the adjoint fermion plays a role of the
center symmetry preserver in the $2^4$-lattice model.
Since the EK volume equivalence is valid for the same lattice
parameters such as the bare coupling and the fermion mass, the
equivalent large-volume theory also has the fermion mass of order
$O(1/a)$ and approximates the quenched theory.
As a comparison, therefore, we consider the $\chi$RMT for the quenched
theory.
We restrict the $\chi$RMT predictions to the case of zero-topological
charge.
Accordingly, the configurations yielding exact zero mode(s) are omitted
from the analysis.

The adjoint QCD with any number of flavors belongs to the universal
class of the Chiral Gaussian Sympletic Ensemble (ChGSE). However, we
also consider two other universal classes, Chiral Gaussian Orthogonal
Ensemble (ChGOE) and Chiral Gaussian Unitary Ensemble (ChGUE), in order
to make the comparison manifest.
The distributions of the lowest eigenvalue are
\beq
P(z)=\begin{cases}
\frac{2+z}{4} e^{-(z/2)-(z^2/8)} & \text{for ChGOE}\\
\frac{z}{2} e^{-z^2/4} & \text{for ChGUE}\\
\sqrt{\frac{\pi}{2}}z^{(3/2)} I_{3/2}(z) e^{-z^2/2} & \text{for ChGSE}
\end{cases},
\eeq
and the spectral densities \cite{Toublan:1999hi} are
\beq
\rho(z)=\begin{cases}
\frac{z}{2}\left[J_0^2(z)+J_1^2(z)\right]+\frac{1}{2}J_0(z)
\left[1-\int_0^z \text{dt} J_0(t)\right] &\text{for ChGOE}\\
\frac{z}{2}\left[J_0^2(z)+J_1^2(z)\right] & \text{for ChGUE}\\
z\left[J_0^2(2z)+J_1^2(2z)\right]-\frac{1}{2}J_0(2z)\int_0^{2z}\text{dt}J_0(t) & \text{for ChGSE}
\end{cases}.
\eeq

Histograms of the lowest twelve Dirac eigenvalues for $N_c=8, 16$ at
$b=0.5$ and $\kappa=0.09$ are shown in \Fig{Dirac_spectrum}.
For the lowest eigenvalue, we found that its distribution is well
described by the $\chi$RMT for the ChGSE (solid blue curve) after
introducing a rescaled eigenvalue $z$ by $z=\lambda V \Sigma$ to fit the
data, where $V=2^4$ and $\Sigma$ is a free parameter.
As a comparison, we show the distribution of the lowest eigenvalue
predicted by the $\chi$RMT for the ChGUE (solid red curve) and ChGOE
(solid black curve) using the same parameter $z$ for the ChGSE.
In addition, we plot the spectral densities for the ChGSE (dashed blue
curve), which are in good agreement with a few lowest eigenvalues.
According to \Fig{web_of_equivalences}, this result would be a strong
evidence that chiral symmetry of quenched large $N_c$ gauge theory
is spontaneously broken.

\begin{figure}
\begin{center}
\includegraphics[width=.5\textwidth]{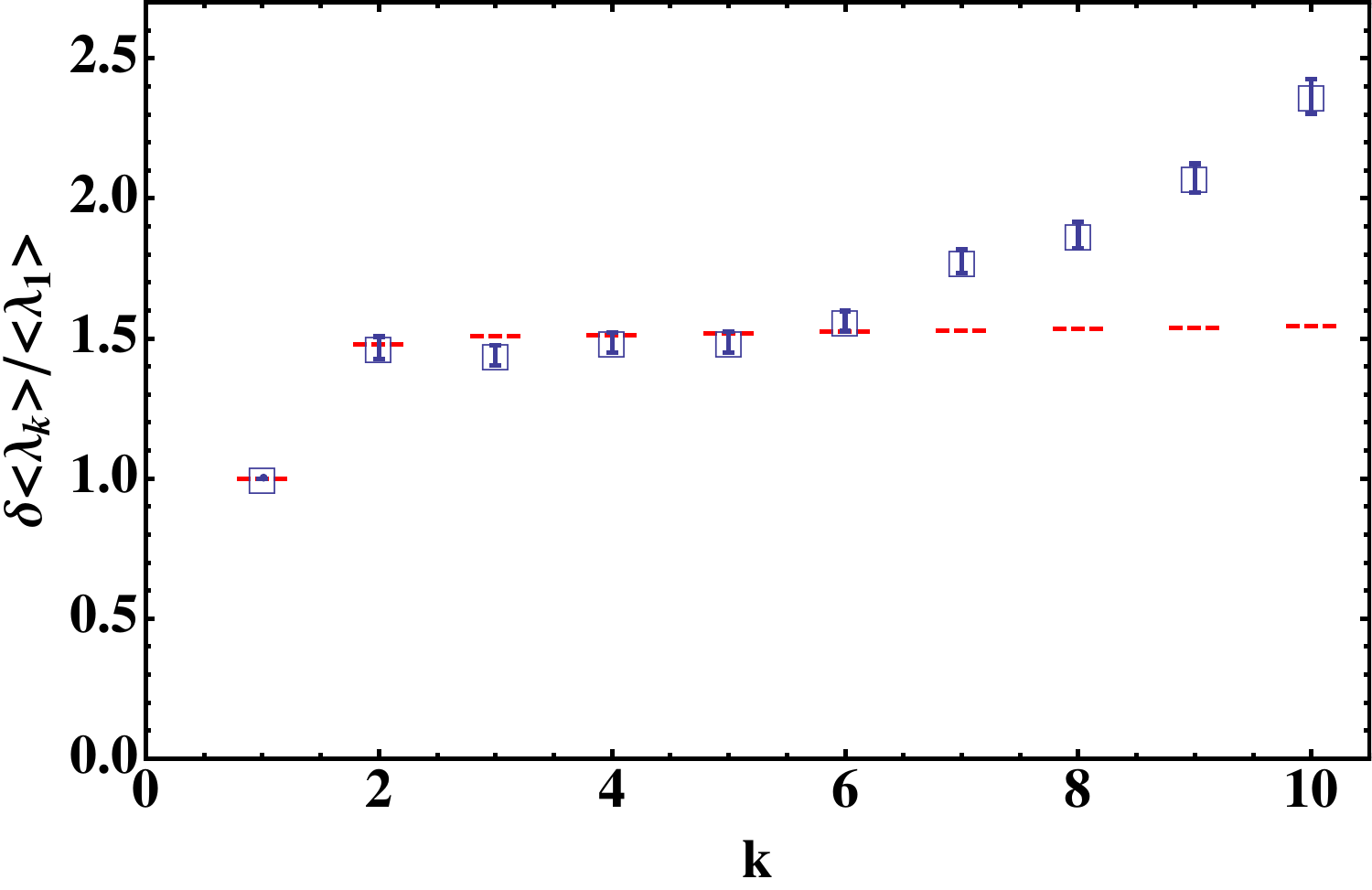}
\caption{%
\label{fig:ratio_delta_lambda}%
 Spacing between the adjacent Dirac eigenvalues normalized by the lowest
 eigenvalue $\langle \lambda_1 \rangle$ for $N_c=16$, where
 $\delta \langle \lambda_k \rangle
 = \langle \lambda_{k} \rangle - \langle \lambda_{k-1} \rangle$ and
 $\langle \lambda_0 \rangle=0$.
 The red dashed lines represent the $\chi$RMT prediction for the
 $\beta=4$ universal class.
}
\end{center}
\end{figure}

In \Fig{ratio_delta_lambda}, we compare $\delta \langle \lambda_k
\rangle/\langle \lambda_1 \rangle$ for $N_c=16$ (our largest value of
$N_c$) with the $\chi$RMT prediction, where the braket
$\langle\cdots\rangle$ represents the ensemble average.
The low-lying Dirac eigenvalues perfectly agree with the $\chi$RMT
prediction, which adds a further evidence for S$\chi$SB.
We have also studied the strong coupling region, $b=0.2$, and similarly found a good agreement with the $\chi$RMT prediction (see \Fig{delta_lambda_bp2}).  
However, note that this point is actually stronger coupling than the bulk phase transition; this phase is not related to the continuum limit a priori.  

\subsubsection{$N_c$-scaling and $gap$ of the Dirac eigenvalues}
\label{sec:eigen_value_spacing}

In \Sec{LargeNcVsRMT}, we argued that the $\chi$RMT limit of the
$2^4$-lattice model is analogous to that of ordinary QCD by replacing
$mV$ with $mVN_c^\alpha$, where the exponent $\alpha$ can be determined
so that $N_c^\alpha$ is the number of degrees of freedom important at
low energy, which is proportional to the inverse of the eigenvalue
density around the origin.
Then, because the low-lying eigenvalues scale as $1/N_c$ as we will
see below, we obtain $\alpha=1$.

\begin{figure}
\includegraphics[width=.5\textwidth]{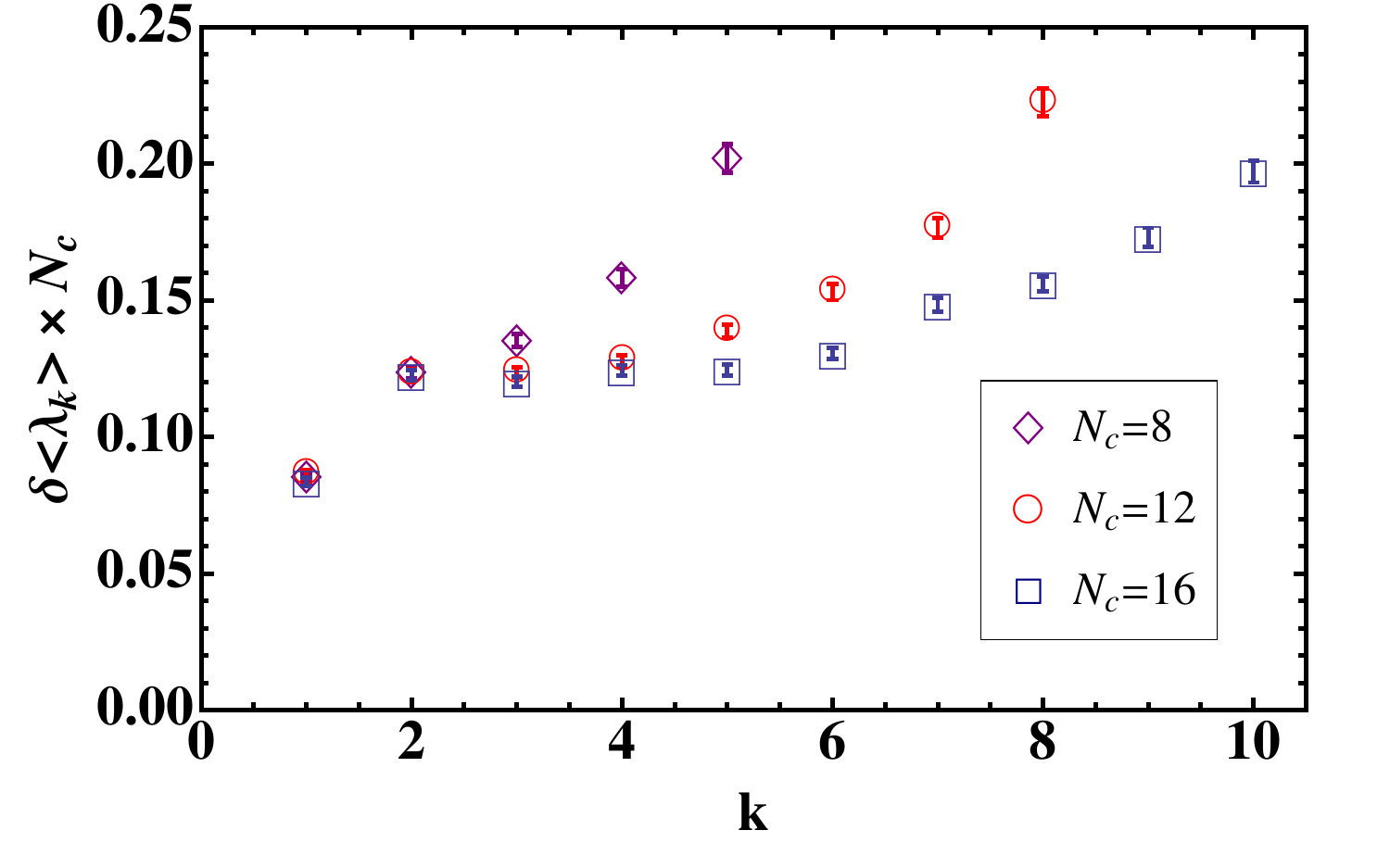}
\includegraphics[width=.5\textwidth]{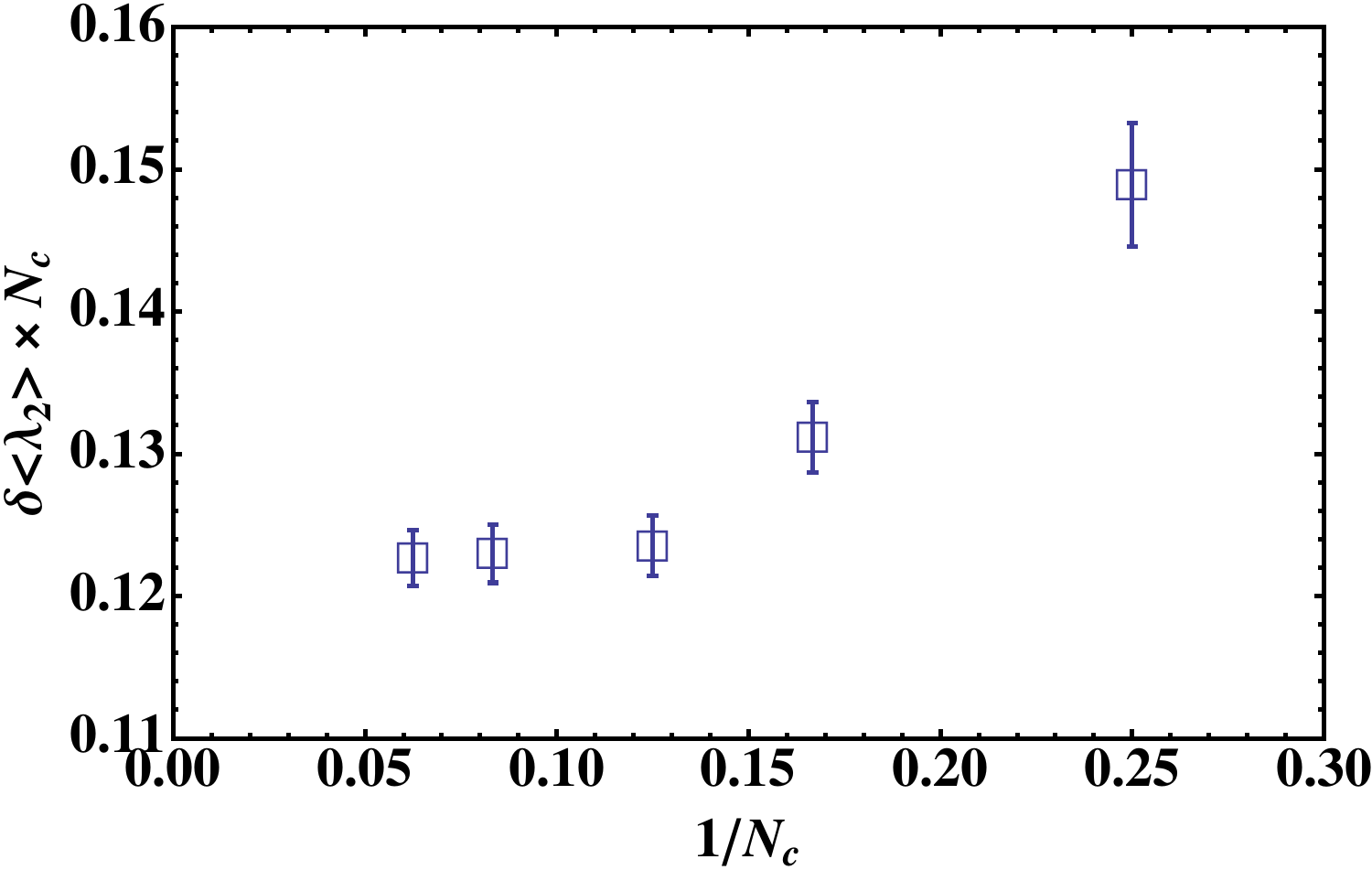}
\caption{%
\label{fig:N_scaling}%
(Left) Spacing between the adjacent Dirac eigenvalues multiplied by $N_c$ for $N_c=8, 12$ and $16$. 
(Right) Spacing between the first and second Dirac eigenvalues multiplied by $N_c$ for $N_c=4,6,8,12$ and $16$.
}
\end{figure}

In the left panel of \Fig{N_scaling}, we plot the spacing between the
adjacent low-lying Dirac eigenvalues multiplied by $N_c$ for $N_c=8, 12$
and $16$.
We see a nice agreement of the data points at up to $k=2$ for $N_c=8$
and at up to $k=4$ for $N_c=12$, where the spacings are expected to show
a plateau for $k\geq 2$.
This result implies that the near-zero eigenvalues, which are expected
to reproduce the $\chi$RMT prediction well, scale as $1/N_c$.
(The same scaling had been reported in \cite{Hietanen:2012ma}, by using
the single-site model with a very light dynamical overlap adjoint
fermion.)
The right figure of \Fig{N_scaling} shows $\delta \langle \lambda_2
\rangle$ multiplied by $N_c$ for various $N_c$; for given statistics,
they agree to each other for $N_c \geq 8$.
We find that the eigenvalue spacing deviates from the $\chi$RMT
prediction as we increase $k$ or decrease $N_c$; the distribution has a
long-tail in the direction of the large eigenvalue.
This behavior can be understood as follows.
As we will see below, the $gap$ appears between the $(N_c-1)$-th and
$N_c$-th eigenvalues, where the spectral density is zero.
The repulsion between eigenvalues becomes weaker as we approach
the $gap$, and thus the distribution develops a long-tail to the $gap$.
Therefore, the eigenvalue spacing near the gap becomes larger than
expected.
Note also that the $1/N_c$ correction takes a rather complicated form
due to the peculiar distribution; for $\delta\lambda_2$, though at
$N_c=4,6,8$ the corerction looks $1/N_c$, at $N_c\ge 8$ this behavior
disappears and the value stays almost constant.
For $\delta\lambda_3$ we observe a similar $1/N_c$-like behavior at
$N_c=8,12$ and $16$, but we expect the  value is saturated at
$N_c\simeq 16$.

\begin{figure}
\begin{center}
\includegraphics[width=.5\textwidth]{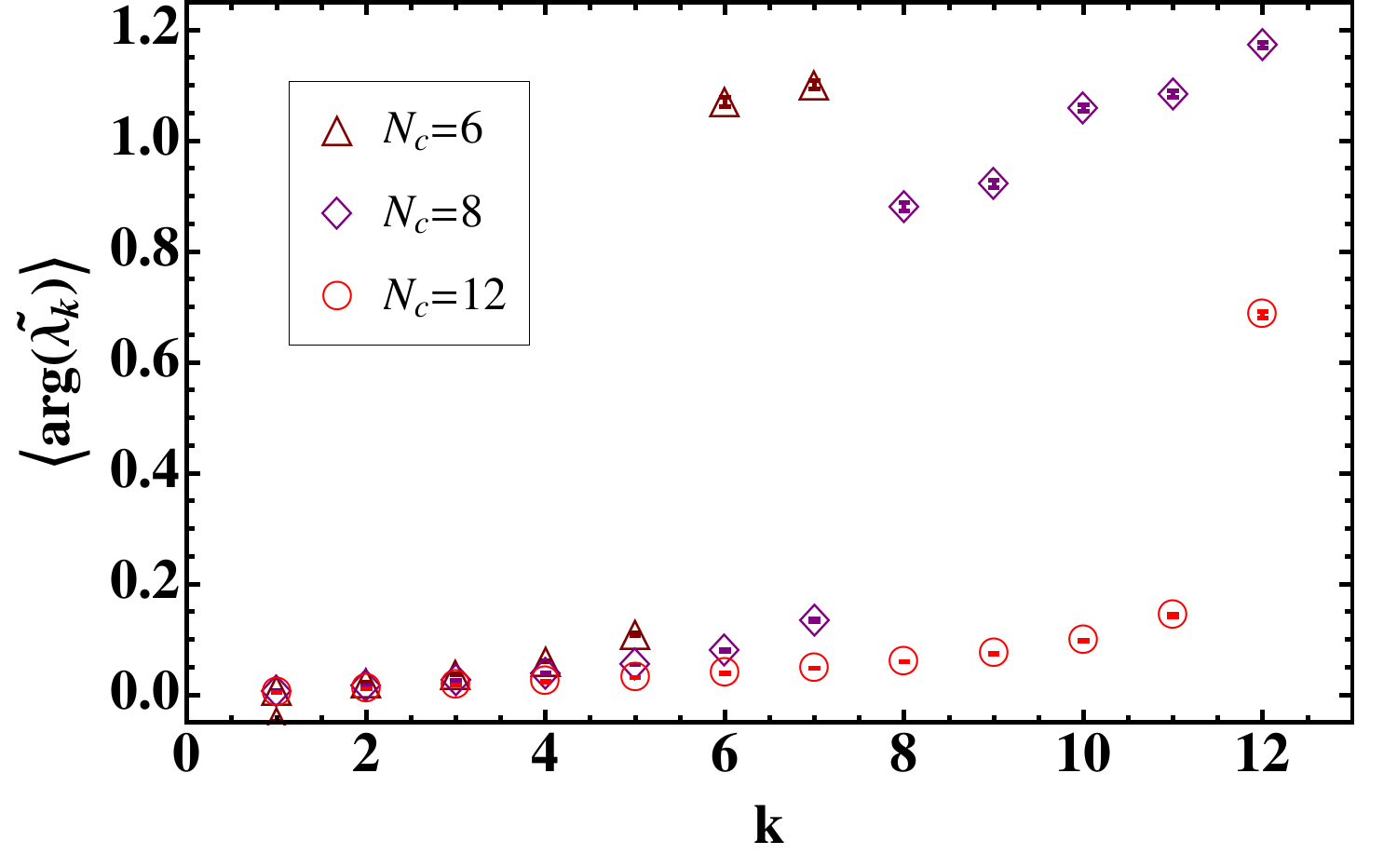}
\caption{%
\label{fig:eigenvalue_gap}%
Low-lying overlap-Dirac eigenvalues in units of radian for $N_c=6, 8, 12$. 
}
\end{center}
\end{figure}

\begin{figure}
\begin{center}
\includegraphics[width=.5\textwidth]{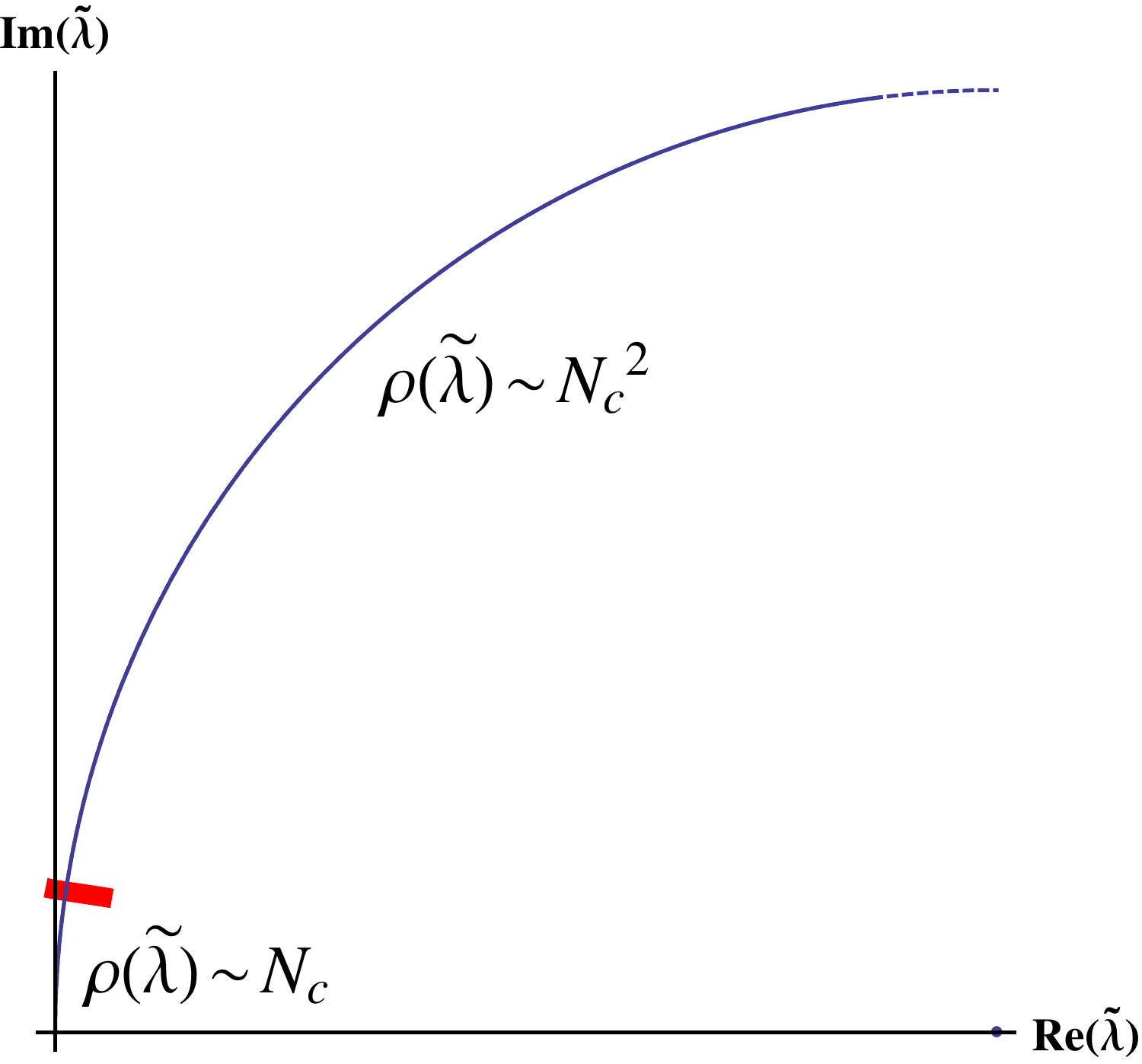}
\caption{%
\label{fig:Dirac_circle}%
(Right) Schematic diagram of the eigenvalue density of the overlap-Dirac operator on a circle in the complex plane. 
}
\end{center}
\end{figure}

To describe this $gap$ clearly, we plot the lowest twelve overlap-Dirac
eigenvalues, which lie on a circle in the complex plain, in units of
radian for $N_c=6,8$ and $12$ in \Fig{eigenvalue_gap}.
The eigenvalue abruptly jumps at $k=5,\ 7$ and $11$ for $N_c=6,\ 8$ and
$12$, respectively.
(A similar $gap$ had also been found in the $N_f=1$ theory
\cite{Hietanen:2012ma}.)
For $k < N_c$, the eigenvalue spacing roughly scales as $1/N_c$ or
equivalently the density scales as $N_c$.
The gap persists in the large-$N_c$ limit: the $N_c$-th eigenvalues are reasonably well  
fit by the function $0.3+4.6 N_c^{-1}$. 
The schematic diagram of the overlap-Dirac eigenvalues on a circle in
the complex plain is shown in \Fig{Dirac_circle}.
The $N_c$-dependence of the eigenvalue density is expected to change
from $N_c$ to $N_c^2$ as $k$ changes from $k<N_c$ to $k\geq N_c$.

A possible explanation of the $1/N_c$-scaling of Dirac eigenvalues and the appearance of a $gap$ relies on the perturbative analysis around the diagonal Wilson lines for a weakly coupled gauge fields. In the perturbation theory, the spectrum of the theory is governed by the background of the Wilson lines and thus the low-lying Dirac spectrum should be closely related to the zero modes \cite{Azeyanagi:2010ne}. 
For the fermions in the adjoint representation the number of zero modes of the Wilson lines is $(N_c-1)$, 
which is the {\it effective} number of degrees of freedom, while the number of total degrees of freedom is $N_c^2$. According to our argument in \Sec{LargeNcVsRMT}, therefore, the low-lying Dirac eigenvalues should scale as $1/N_c$, being consistent with our numerical results. 
We emphasize that this analysis is only true for weakly coupled large $N$ gauge theory in $a~compact~space$. 

When $N_f=1$, the position of this $gap$ has been studied for several
values of the coupling constant \cite{Hietanen:2012ma}, which turned out
to be almost independent of the coupling constant in the lattice
unit.
Therefore, in the physical unit, the scale corresponding to the location
of the gap diverges as the lattice cutoff increases.
This result looks natural, because a new physical scale appears
otherwise.

It is also satisfactory from the universality point of view:
$SU(N_c)$ theory with $N_f$ adjoint fermions and $SO(2N_c)$ theory with
$N_f$ fundamental fermions are equivalent in the $\chi$RMT limit,
because they are described by the same $\chi$RMT.
However they are completely different in the 't Hooft large-$N_c$ limit;
whereas the former has $O(N_c^2)$ fermion degrees of freedom, the latter
has only $O(N_c)$.
In order for them to become identical in the $\chi$RMT limit, the
fermionic degrees of freedom must match somehow.
But now we know the mechanism: only $O(N_c)$ degrees of freedom survives
in the low energy limit of  the adjoint theory, because $N_c^2-N_c$
eigenvalues become infinitely large.
(In the case of the fundamental fermions in the probe limit, there is no
$gap$ \cite{Narayanan:2004cp}).

\begin{figure}
\includegraphics[width=.5\textwidth]{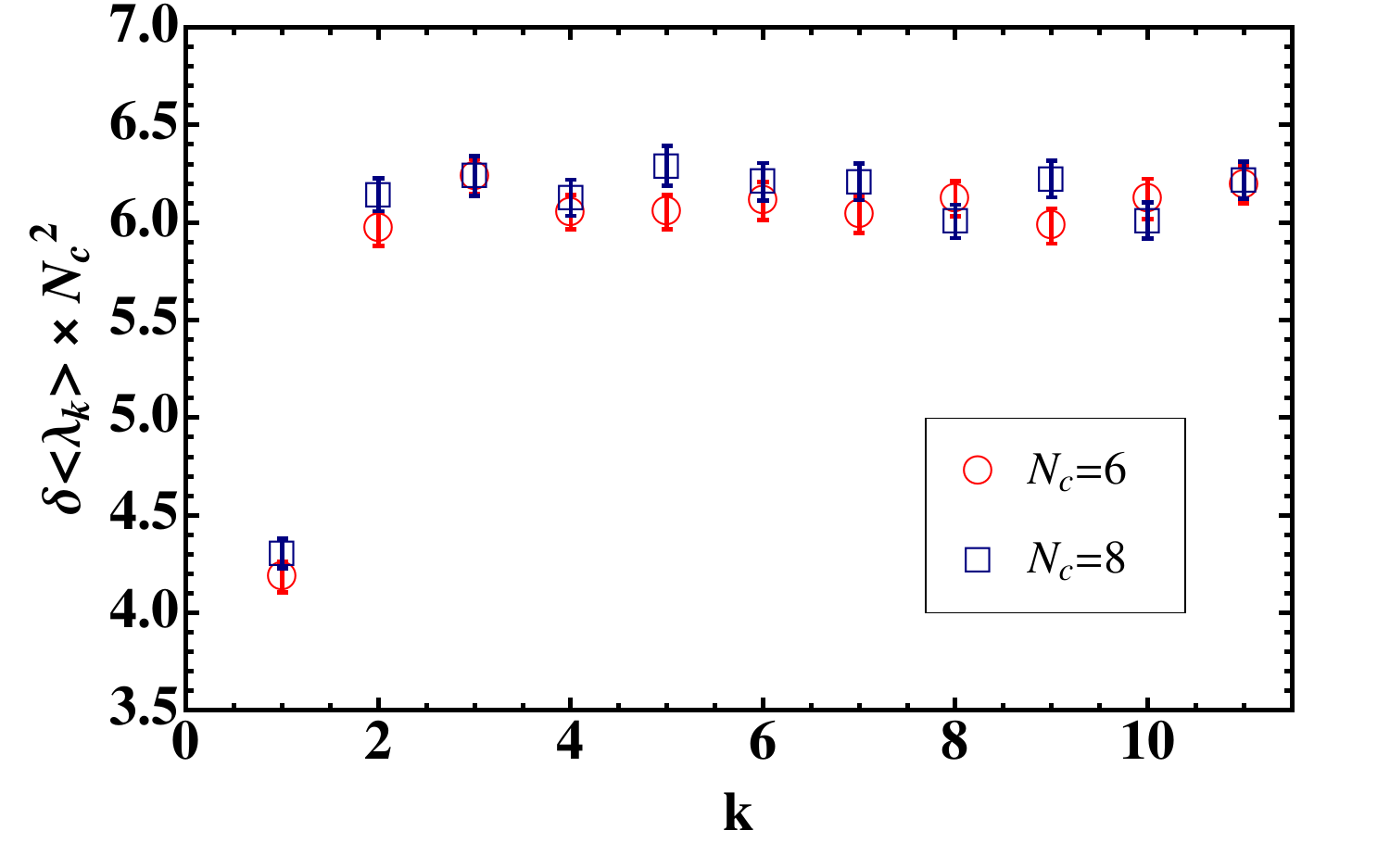}
\includegraphics[width=.5\textwidth]{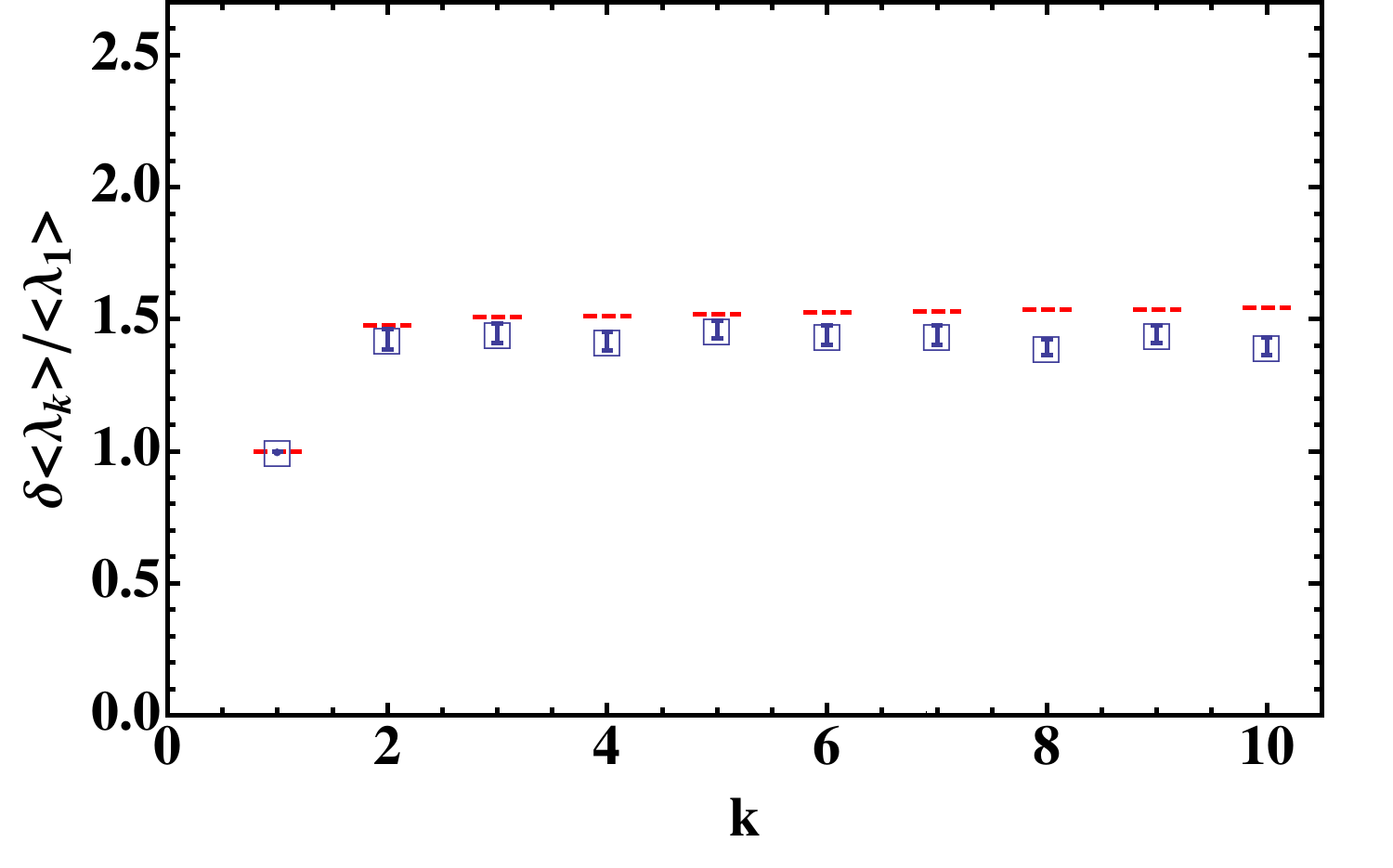}
\caption{%
\label{fig:delta_lambda_bp2}%
(Left) Spacing of the adjacent Dirac eigenvalues multiplied by $N_c^2$ for $N_c=6, 8$ with $b=0.2$, where $\delta \langle \lambda_k \rangle=\langle \lambda_{k}\rangle-\langle \lambda_{k-1} \rangle$ and $\langle \lambda_0 \rangle=0$. 
(Right) Spacing between the adjacent Dirac eigenvalues normalized by the lowest eigenvalue $\langle \lambda_1 \rangle$. The red dashed lines represent the $\chi$RMT prediction for the $\beta=4$ universal class.
}
\end{figure}

At strong coupling phase, the perturbative treatment around the diagonal
background Wilson lines is no longer reliable: the zero modes can be
lifted by gauge fluctuation and the off-diagonal components would be the
same order of magnitude of the diagonal components.
In contrast to the case of weak coupling, therefore, we expect that the
$gap$ is absent and the eigenvalue spacing is of order $O(1/N_c^2)$.
In the left panel of \Fig{delta_lambda_bp2}, we plot the eigenvalue
spacing multiplied by $N_c^2$ for $N_c=6$ (red circle) and $N_c=8$ (blue
square) and at $b=0.2$.
The data shows a nice plateau and agrees to each other, implying that
the eigenvalue scales as $1/N_c^2$ without any $gap$ in the strong
coupling regime, as expected.
In the same manner, it is expected that the Dirac eigenvalues scale as
$1/N_c^2$ and the gap does not exist even at weak coupling if the volume
is sufficiently large: as the volume increases the momentum gap between
the lowest and the first excited state, which differ by ($2\pi/L$),
decreases and 
at some point the zero modes are lifted by gauge fluctuation.


\section{Conclusion and discussion}
\label{sec:conclusion}

In this paper we considered how to apply the $\chi$RMT techniques to large-$N_c$ gauge theories. 
After giving general considerations, we provided a numerical demonstration by using the $2^4$-lattice model as an example. 
The most important lesson is that the 't Hooft large-$N_c$ limit and the $\chi$RMT limit (the microscopic limit) are not compatible in general: 
the former is the large-$N_c$ limit with the fermion mass $m$ and the space-time volume $V$ fixed, 
while the latter requires $mVN_c^\alpha$ to be fixed, where $\alpha$ is a positive constant which may depend on the theory.  
The value of $\alpha$ is a unity in the example we studied in Sec. 5, which is different from the usual 't Hooft counting. 

An important consequence of the difference between two limits is that 
several properties in the 't Hooft large-$N_c$ limit (e.g. the equivalence between $SU(N_c)$, 
$SO(2N_c)$ and $USp(2N_c)$ gauge theories) do not hold in the $\chi$RMT limit. 
This fact must be appreciated when one applies the large-$N_c$ and/or $\chi$RMT approaches to QCD and related theories;  
although these two approaches provide valuable `exact' results, they are valid in different parameter regions and hence 
one has to carefully choose more suitable method depending on the physics he/she studies. 
In spite of the difference of two limits,  
one can still detect S$\chi$SB of large-$N_c$ gauge theories 
as we have demonstrated in Sec.~\ref{sec:numerical_result}. 

Rather curiously, we observed a nice agreement between the $\chi$RMT and $2^4$-lattice model 
even when the center symmetry in the latter is broken spontaneously. Of course we cannot relate this fact to S$\chi$SB 
in the large-volume theory, because neither the EK equivalence nor the analytic continuation can be used. 
This is presumably because the space-time dimension is not important for the universality argument; only the pattern of S$\chi$SB matters.

As a next step, we are studying whether the $SU(N_c)$ gauge theory with dynamical adjoint fermions  
goes through S$\chi$SB or not, with an application to technicolor models in mind. 
We hope to report the result in near future.

\section*{Acknowledgements}
\hspace{0.51cm}
The authors would like to thank S.~Hashimoto, A.~Hietanen, M.~Kieburg, M.~Koren, S.~Sharpe, M.~Tezuka, J.~Verbaarschot and N.~Yamamoto for stimulating discussions and comments. 
The numerical computations used in this work were carried out on cluster at KEK.  
M.~H. would like to thank Boston University for hospitality at the final stage of this work. 
This work is supported in part by the Grant-in-Aid for Scientific Research of the Japanese Ministry of Education, Culture, Sports, 
Science and Technology and JSPS (Nos. 20105002,20105005,and 22740183).

\appendix

\bibliographystyle{ieeetr}
\bibliography{AdjQCD_largeN}

\end{document}